# Social-Sensor Composition for Tapestry Scenes

Tooba Aamir, Hai Dong, *Member, IEEE*, and Athman Bouguettaya, *Fellow, IEEE*

**Abstract**—The extensive use of social media platforms and overwhelming amounts of imagery data creates unique opportunities for sensing, gathering and sharing information about events. One of its potential applications is to leverage *crowdsourced* social media images to create a tapestry scene for scene analysis of designated locations and time intervals. The existing attempts however ignore the temporal-semantic relevance and spatio-temporal evolution of the images and direction-oriented scene reconstruction. We propose a novel social-sensor cloud (SocSen) service composition approach to form tapestry scenes for scene analysis. The novelty lies in utilising images and image meta-information to bypass expensive traditional image processing techniques to reconstruct scenes. Metadata, such as geolocation, time, and angle of view of an image are modelled as non-functional attributes of a SocSen service. Our major contribution lies on proposing a context and direction-aware spatio-temporal clustering and recommendation approach for selecting a set of temporally and semantically similar services to compose the best available SocSen services. Analytical results based on real datasets are presented to demonstrate the performance of the proposed approach.

## 1 Introduction

THE proliferation of smartphones, together with the popularity of cloud-based social networks provides a unique paradigm for data sensing, gathering and sharing [1], [2]. This data sensing and gathering paradigm can be termed as *social sensing* [1]. Social-sensing allows crowdsourcing massive amounts of multi-modal data from multiple *social-sensors* [1]. The multi-modal data (i.e., *social-sensor data*) can be provisioned via smartphones (i.e., social-sensors) and hosted in social networks (also termed as *social clouds* [2] or *social-sensor clouds* [3], [4]). Examples of social-sensor data include Twitter posts and Facebook status. Social-sensor data has become a significant and readily available means for expressing opinions about activities and events. Posts related to public events or incidents, especially images and related descriptions, may contain critical information [1]. Traditional sensors like CCTV usually do not provide such comprehensive coverage. The ubiquity of smartphones with high-resolution cameras enables the sharing of images any time and anywhere. As a result, whenever an event happens, it is quite likely to be photographed by one or more smartphone users. Such photographs can provide a *spatio-temporal view* of the event [1], [4].

The traditional approaches of event scene reconstruction or analysis rely on data mining and image processing [9], [10], [19], [20], [22], [23]. Most of the current work focuses on utilising and improving the performance of scene capturing hardware and analysis software. These techniques are both costly and time-consuming [19], [20]. Social-sensor data allows *low cost*, *fast* and *accessible coverage* of spatio-temporal events. Therefore, social-sensor data is considered as a promising and emerging source for monitoring and analysing events. However, there are numerous inherent challenges regarding management and efficient delivery of the social-sensor data. The social-sensor data is collected from multiple sources in several formats. It is challenging to *manage the multifaceted social-sensor data* in the interoperable social cloud platforms. Another significant challenge is *searching and analysing the massive amount of heterogeneous data* without exposing the complexities of data collection and management.

*Service paradigm* can address the two challenges. Abstracting a social media image as a service helps to manage and transform social-sensor data into useful information [3], [4]. Service paradigm abstracts social-sensor data streams into *small independent functions*, i.e., *social-sensor cloud services* (abbreviated as SocSen services). These services have multiple *functional* and *non-functional attributes*. Utilising social-sensor data as services add value to the unstructured social-sensor data and results in uniform and ubiquitous delivery. The functional attributes include shooting mode, descriptions, tags, etc. The spatio-temporal and contextual information of a social-sensor image can be presented as the *non-functional attributes* of a service. The qualitative features (e.g., price, trust, etc.) are conceptualized as the *non-functional attributes* of the service. Abstracting social-sensor data into services reduces the *complexity* of social-sensor data and supports *efficient and real-time access* to high-quality and relevant data. SocSen services also make it easy to access and reuse social-sensor data in multiple applications, over different platforms.

This paper focuses on using service paradigm to devise an approach to reconstruct a scene, *without implementing image processing*. We specifically focus on the images and the image meta-data downloaded from social clouds to monitor

- *T. Aamir and H. Dong are with the School of Science, RMIT University, Melbourne, VIC 3000, Australia.*
*E-mail: {tooba.aamir, hai.dong}@rmit.edu.au.*
- *A. Bouguettaya is with the School of Computer Science, University of Sydney, Sydney, NSW 2006, Australia.*
*E-mail: athman.bouguettaya@sydney.edu.au.*

*event-specific scenes*. We propose *a new approach for selecting, processing and analysing social-sensor cloud data as services* to reconstruct and analyse the spatio-temporal events. A complete scene analysis needs to assemble context-relevant images obtained from *multiple directions (i.e., the point of view from the camera location), locations*, and different time intervals. Initial attempts have been made to address these requirements [4], [5]. However, we have identified the following open issues:

- *Temporal-semantic relevance and spatio-temporal evolution:* Temporally and semantically similar spatio-temporal services help to identify an event's activity patterns in space and time. he spatio-temporal evolution of an event facilitated by the temporal-semantic similarity can assist in generating the event's timeline. Less attention in existing studies has been given to these two aspects.
- *Direction-oriented spatio-temporal composition:* The various directions of service coverage is a challenge for the SocSen service composition. A successful composite service should be able to depict the whole scene from a specific direction or point of view of coverage. No efforts in previous research have been made towards this issue.

This paper accommodates the solution of the aforementioned issues and proposes *a novel recommendation approach for SocSen service composition*. The proposed recommendation approach helps to discover the direction-aware spatio-temporally relevant services for SocSen service composition with minimum human intervention. The recommendation approach has made two major contributions.

1) The approach proposes *a model for selecting a set of temporally and semantically similar services*. The selection process recommends a set of clusters of SocSen services from spatial, temporal and semantic perspectives. The given set of semantically similar services exists close in time, but might be spatially far. The semantically similar spatio-temporal service clusters help to make up an event's activity pattern.
2) The approach proposes *a context and direction-aware spatio-temporal clustering approach*. The proposed approach of clustering helps to compose the SocSen services considering the directions of the service coverage. The proposed approach of clustering helps to handle multiple queries with similar nature. The new composition approach is based on the vantage point of service coverage.

The proposed SocSen service composition forms a *tapestry scene in the spatial aspect* and a *storyboard in the temporal aspect*. In the spatial aspect, the composition forms *a tapestry scene* by selecting individual relevant services. In the temporal aspect, an event *timeline* is formed by combining various *individual tapestries* and placing them in a *bigger temporal tapestry* to form the story of an event. The individual tapestries are the tapestries formed by composing the services in individual clusters. The composite service aims to provide *a user-required view and related information* about an event for the scene analysis.

The rest of the paper is structured as follow. Section 2 reviews the related background work. Section 3 describes the motivating scenario. Section 4 formally defines the SocSen service. Section 5 explains the SocSen service composition approach for a tapestry scene. Section 6 describes the evaluation of the proposed approach. Section 7 concludes the proposed work.

## 2 Related Work

Our SocSen service selection and composition approach is built upon existing work in scene analysis, social-sensing, sensing-as-a-service, and service selection and composition.

- *Scene analysis and tracking* are active research areas for the development of robust surveillance and monitoring systems [9], [10], [25]. Event monitoring or scene analysis through social network feeds is still new. Visual concept detectors are applied on camera video frames and interpreted as camera tweets posted regularly [22]. These tweets are represented by a combined probabilistic spatio-temporal (PST) data structure which is then combined with the concept-based image (Cmage) as the common representation for visualisation. Topic models and graph-based ranking are used to propose a multi-modal event summarisation framework for social media in [23]. This summarisation framework uses a set of social media posts about an event and selects a subset of shared images by maximising their relevance and minimising their visual redundancy.

- *Social sensing* assists in almost real-time visualisation of life events via photos, blogs, etc. shared on various social media sites. Images and videos are especially crucial because they can be used to build a visual summary of an event and further used in other applications, e.g., area monitoring and scene analysis. Spatio-temporal social media analysis for event detection is discussed in various studies including traffic planning [6], environment monitoring [13], [34] and so on [11], [12]. Abnormal topics, individuals and events within various social media data sources for visual analysis are examined in [24], [32]. Another study proposes an approach towards multi-scale event detection using social media data. It takes into account different temporal and spatial scales of events in the data [9]. A location-based event detection approach using locality-sensitive hashing is proposed in [10]. This approach is used to detect real—-world events by clustering micro-blogs with high similarities. Researchers also proposed a framework to detect composite social events over streams using a graphical model called location-time constrained topic (LTT) [26]. This framework fully exploits the information about social data over multiple dimensions. Most of these approaches are *data-centric*, built upon *data mining and analysis techniques*. Data mining and analysis techniques require *a considerable amount of expertise and time*. Accessing and analysing the social-sensor data obtained from various social clouds results in several shortfalls like, *the incapability of the time efficient and smooth processing of this data*. SOA-based solutions only consider spatio-temporal and contextual aspects of sensor data for scene analysis without exploiting the data complexities. Thus, using SOA and social-sensors for scene analysis is believed to be more efficient than using image processing over the batch of images to build the scene.

- *Crowdsourcing and sensing-as-a-service* are large-scale sensing paradigms based on the power of IoT devices, including mobile phones, smart vehicles, wearable devices, and so on [15], [18]. Crowdsourcing allows the increasing number of mobile phone users to share local knowledge (e.g., local information, ambient context, noise level, and traffic conditions) acquired by their sensor-enhanced devices. The information can be further aggregated in the cloud for large-scale sensing and community intelligence mining [12]. The mobility of large-scale mobile users makes sensing-as-a-service a versatile platform that can often replace static sensing infrastructures [15], [30]. A broad range of applications is thus enabled, including traffic planning [6], environment monitoring and public safety [25], [33], mobile social recommendation [24], and so on. From the AI perspective, sensing-as-a-service is founded on a distributed problem-solving model [27].

- *Service selection and composition* are the major research problems in service-oriented computing [8], [17]. Service selection and composition have been applied in many domains, including scene analysis and visual surveillance [5], [28]. The service composition problem can be categorised into two areas. The first area focuses on the functional composability among component services. The second area aims to perform service composition based on non-functional attributes (QoS). In [29], service composition from media service perspective has been discussed. In [35], service skyline model is introduced for selection of optimal sets of services as an integrated service package over multiple sets of services. A mashup based approach to utilise existing online services and data into entirely new applications is investigated in [18]. The proposed approach enables users to create, use, and manage mashups with little or no programming effort. A composition approach for the sensor clouds and crowdsourced services based on the dynamic features such as spatio-temporal aspects is proposed in [6] and [11]. In our previous work, we proposed a conceptual model for SocSen services as an open and flexible solution for scene monitoring applications [3], [4], [5]. SocSen service selection and composition are proposed for spatio-temporal scene analysis in [4], [5]. However, SocSen service composition for a scene from *specific coverage and context perspectives* is yet to be explored.

## 3 Motivation Scenario

We use an unsettled traffic movement scenario as our motivating scenario. Let us assume, members of the public unite want to gather intelligence on erratic drivers in Town 'ABC'. Some residents posted about an *erratic car driving* on *20/01/17* around *13:30*, at the *intersection of Road X and Road Y*. Posts and images report a *"red car mounts footpath"* and *"accident"* occurred *"around 1:45 pm"*, along *"Road Y"*, as shown in Fig. 1. Traffic command operations want to analyse and reconstruct the car accident *timeline*. They want to find the cause and effect of the incident. Their objective is to *find the behaviour leading to the accident and the vehicles or people involved*.

We propose to *leverage these social media images to reconstruct the desired scene that narrates the complete story for the traffic command operations to perform scene analysis*. Our goal is *a spatio-temporal and contextual coverage of the chain of events resulting in the accident and objects of interest involved*. The query $q = < A, t_s, t_e, d >$ includes an approximate region of interest, an initial textual description of the queried scene, and the approximate time frame of the queried event.

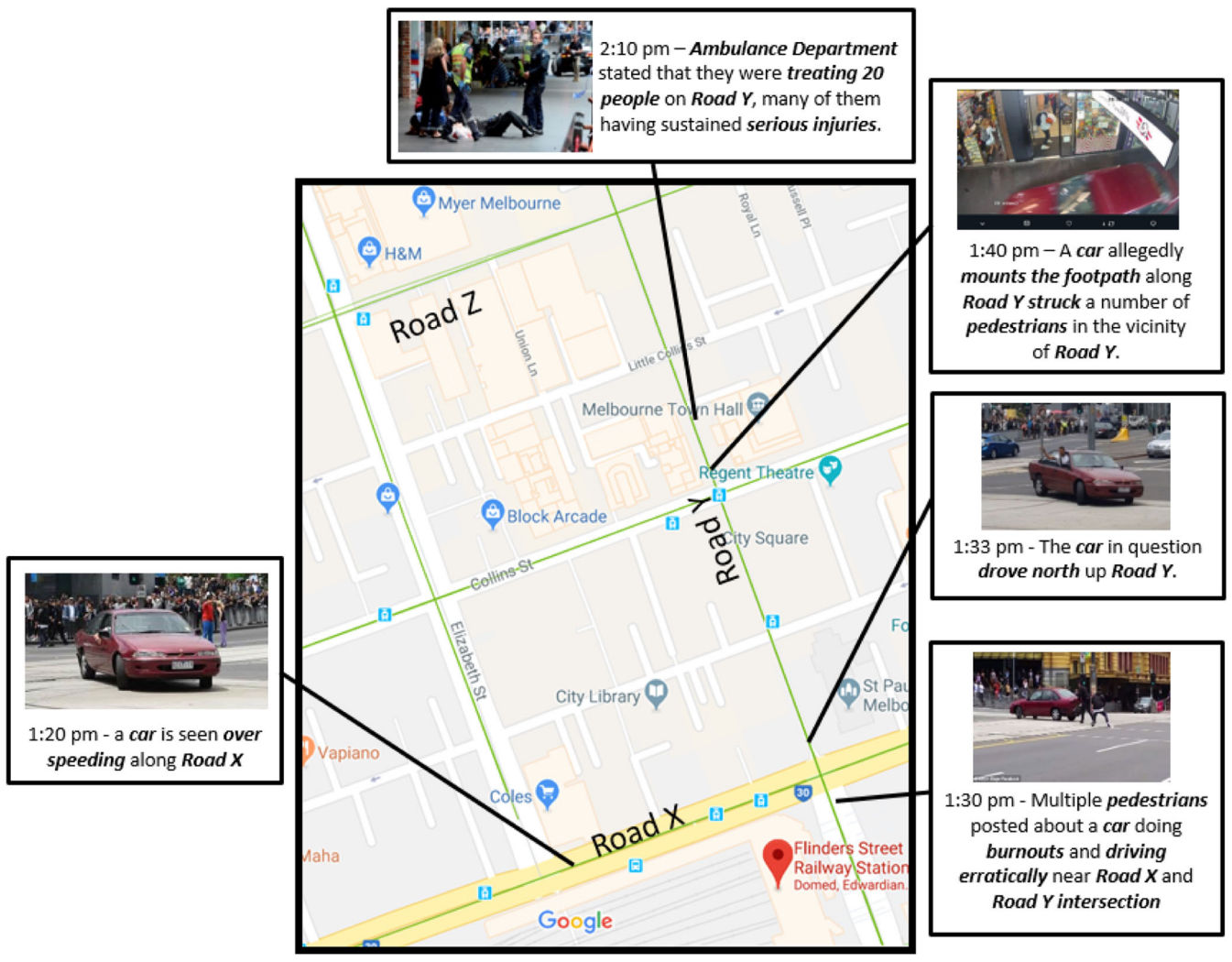

Fig. 1. Motivation scenario - incidents.

- Query Phrases $d$, a set of phrases describing the query, e.g., 'accident', 'red car', 'intersection Road X and Y', 'construct timeline', 'cause of accident', 'effect of accident'
- Query Region $A(< x, y >, l, w)$, where $< x, y >$ is a geospatial co-ordinate set, i.e., decimal longitude-latitude position and l and w are the horizontal and vertical distances from the center $< x, y >$ to the edge of region of interest, e.g., (-37.8101008,144. 9634339, 10 m, 10 m)
- Query Time $(t_s, t_e)$, where $t_s$ is the start time of the query, e.g., 13:20 AEST, 20/01/2017 and $t_e$ is the end time of the query, e.g., 14:00 AEST, 20/01/17.

The process of providing investigators with extra visual coverage can be considered as *a service selection and composition problem*. This work proposes a model for selection and composition of SocSen services based on the user query. SocSen services have multiple functional and non-functional attributes. The basic functional attributes of a service include 1) Time $T$ of the service at which the image is taken and 2) A set of keywords or key-phrases $M$ providing additional information regarding the service, e.g., 'road accident' and 'accident on Road X' etc. The basic non-functional attributes include 1) Service location $L(x, y)$, i.e., longitude and latitude positions of the service and 2) Coverage $Cov$ of the service defined using the maximum visible distance covered by the service, the direction of the service and the angular extent of a scene covered by the service.

In the proposed scenario, the accident involves a single vehicle. Car A mounted the pavement and hit pedestrians at the intersection of Roads X and Y (Fig. 1). The aim is to *reconstruct a visual summary of the accident scene*. The visual summary is constructed by finding the available atomic services and forming a composite service. The proposed composite service is constructed using the *functional and*

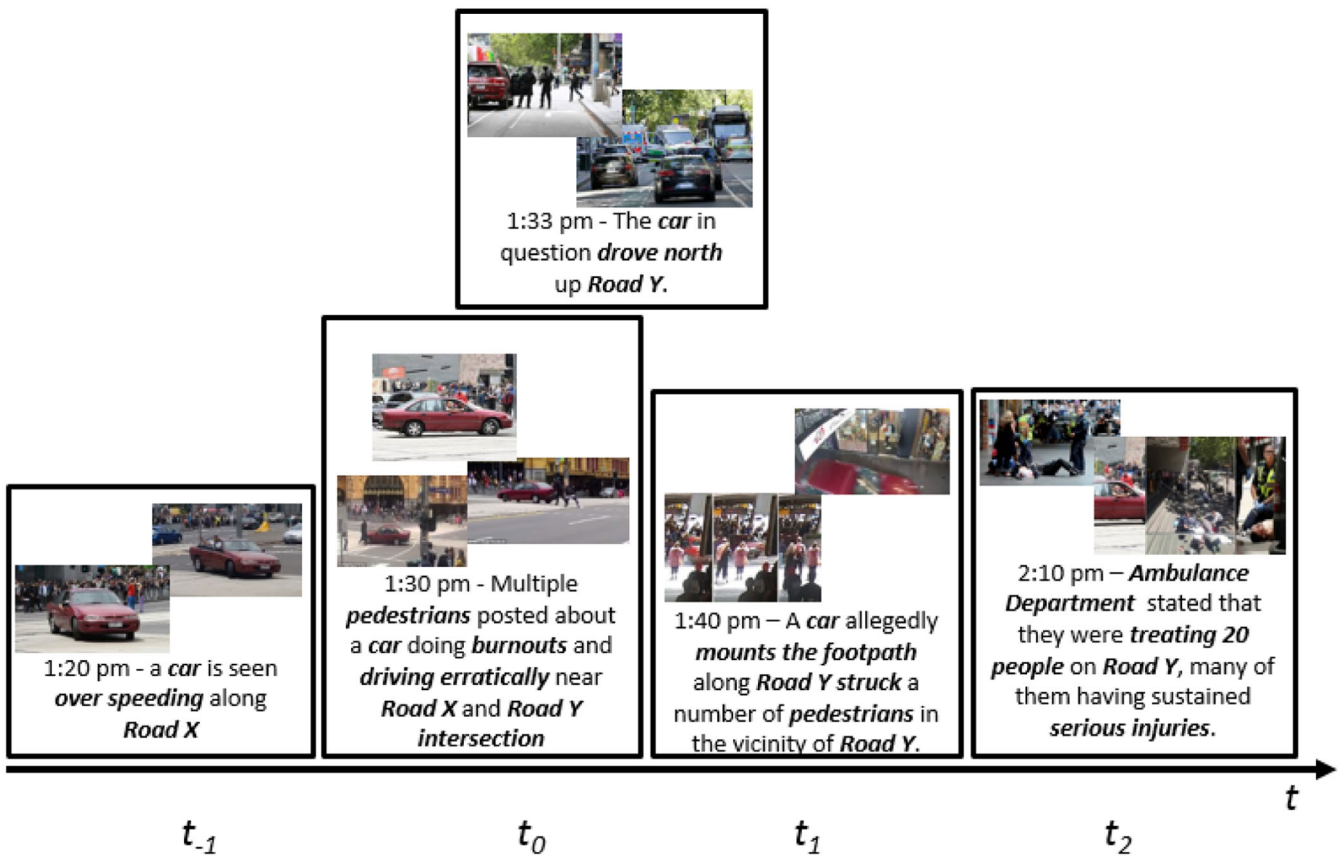

Fig. 2. Motivation scenario - incident timeline.

*non-functional attributes* of the atomic services. The visual and textual summary captures the *involved objects' behaviour* leading to and after the specified event.

It is assumed that all available services are associated with *a two-dimensional geo-tagged location and time*. All the available services are *indexed* for location $L(x, y)$ of query $q$. Indexing is achieved by considering their spatio-temporal features using a 3D R-tree [4]. The search space is reduced by *clustering* and *selecting* the spatio-temporally close and contextually related services. The clustering is based on spatio-temporal features and direction (i.e., the point of view from the camera location) of the coverage. Spatio-temporal and directional clustering helps to handle multiple queries of a similar nature. Semantic context raised from each spatio-temporal service is used for the selection of the contextually related services. Semantic similarity is used to determine contextually related services. *Semantic similarity* is based on the textual similarity between semantic annotations of the services and the query description. A *vocabulary dictionary* of domain experts is used to assess the context-specific textual similarity. For example, the dictionary includes the example behaviours like racing, burnout and tailgating for a query related to *erratic driving*. In addition, the dictionary also includes behaviour synonyms, e.g., reckless and careless. We assess and select services according to such spatio-temporal and contextual relationships.

As shown in Fig. 2, at time $t_{-1}$, i.e., 13:20, two images (i.e., services) show that Car A was racing along Road X. Further, at time $t_0$ (i.e., 13:30) three services show that Car A made a stop at the intersection and was seen driving erratically and endangering the public. Pedestrians in danger, due to erratic behaviour, are considered as interacting with the object of interest, Car A. Two other services show that at the intersection, Car A drove up north on Road Y, at 1:33 pm. Following this at time $t_1$ (i.e., 13:40) two services present that Car A mounts the footpath along Road Y. Further, at time $t_2$ (i.e., 14:10) two services show that people are getting treated by paramedics.

Eleven services are selected to reconstruct the car accident timeline. We *compose* the chosen services according to their spatio-temporal and contextual relationships and composability. Finally, we complete a composition, i.e., the visual summary is created by forming a tapestry scene. The tapestry is formulated by *selecting the composable services covering the accident* and *placing images in the spatial grids* following the path of objects of interest. The composite service is a series of temporal tapestries in space, providing a visual summary of the spatio-temporal evolution of the event. The composition ascertains the uncertain parameters, like *related events, their timeline, other objects involved and their interaction*. For example, overspeeding on Road X is related to the accident on the Road Y and pedestrians affected by this accident are also objects of interest. As shown in Fig. 2, the tapestry manifests the incident timeline, involved vehicle, and other related factors.

## 4 Model for Social-Sensor Cloud Services

The main concepts related to the SocSen service include:

**Definition 1.** *Visual summary* VisSum *is defined as a set of 2D images covering the scene of an event. This observation is a tapestry of context-relevant images in similar spatial and temporal dimensions. A* VisSum *gives users an accurate impression of what a particular scene looks like. An optimal* VisSum *captures the* involved objects' behaviour *leading to and after the specified event. An optimal* VisSum *is constructed by finding and selecting the most appropriate atomic services and forming a composite service. The proposed composite service is constructed based on appropriate composability parameters of spatio-temporal and textual similarity.*

**Definition 2.** *Social-sensor* SocSen *is a camera device capturing an image shared on social media. A social-sensor provides information like camera GPS location, viewing direction and distance, and time when the image was captured.*

**Definition 3.** *Crowd-sourced social-sensor owner* SocSenOwn *is a user of the social media, who provides a service. A social-sensor owner provides contextual information about the service. The contextual information includes a textual description of an image specifying information related to a scene captured, e.g., objects, places, and actions captured in the image.*

**Definition 4.** *Social-sensor cloud* SocSenCl *is a social media platform hosting data from the social-sensors. A social-sensor cloud provides interaction information, i.e., the details and descriptions of relevant services. Such relevant services might cover objects interacting with objects of interest and details of their interactions. It is defined by:*

- *Social-sensor cloud ID* SocSenCl_id.
- *Sensor set* SenSet = $\{SocSen_id_i, 1 \leq i \leq m\}$ *represents a finite set of sensors* SocSen *that collects and hosts sensor data in the respective cloud.*

Table 1 illustrates a summary of all the notations used in this paper.

### 4.1 Model for an Atomic Social-Sensor Cloud Service

Atomic SocSen service *Serv* is defined by:

- *Serv_id* is a unique ID of the service.
- *SocSenCl_id* is the cloud ID where the service locates.
- *F* is a set of functional attributes.
- *nF* is a set of non-functional attributes.

TABLE 1
Summary of Notations

| Notation | Definition |
|---|---|
| *VisSum* | Visual summary |
| *SocSen* | Crowd-sourced social-sensor |
| *SocSenOwn* | Crowd-sourced social-sensor owner |
| *SocSenCl* | Social-sensor cloud |
| $q$ | User's query |
| *q.Rgn* | Query region $\{P < x, y >, l, w\}$ |
| *Serv* | Atomic social-sensor cloud service |
| *L(x,y)* | *Serv*'s geo-tagged location |
| $t$ | Time of the *Serv* or $q$ |
| $t_s$ | *Serv*'s start time |
| $t_e$ | *Serv*'s end time |
| *Cov* | Visual coverage of the *Serv* |
| *VisD* | The maximum visible coverage distance of the *Serv* |
| $\overrightarrow{dir}$ | The visual direction of the *Serv* |
| $\alpha$ | The angular extent of the scene covered in the *Serv* |
| *Con* | Contextual description of the *Serv* |
| *D* | Descriptive information regarding the image of the *Serv* |
| *T* | Location and focus of the image of the *Serv* |

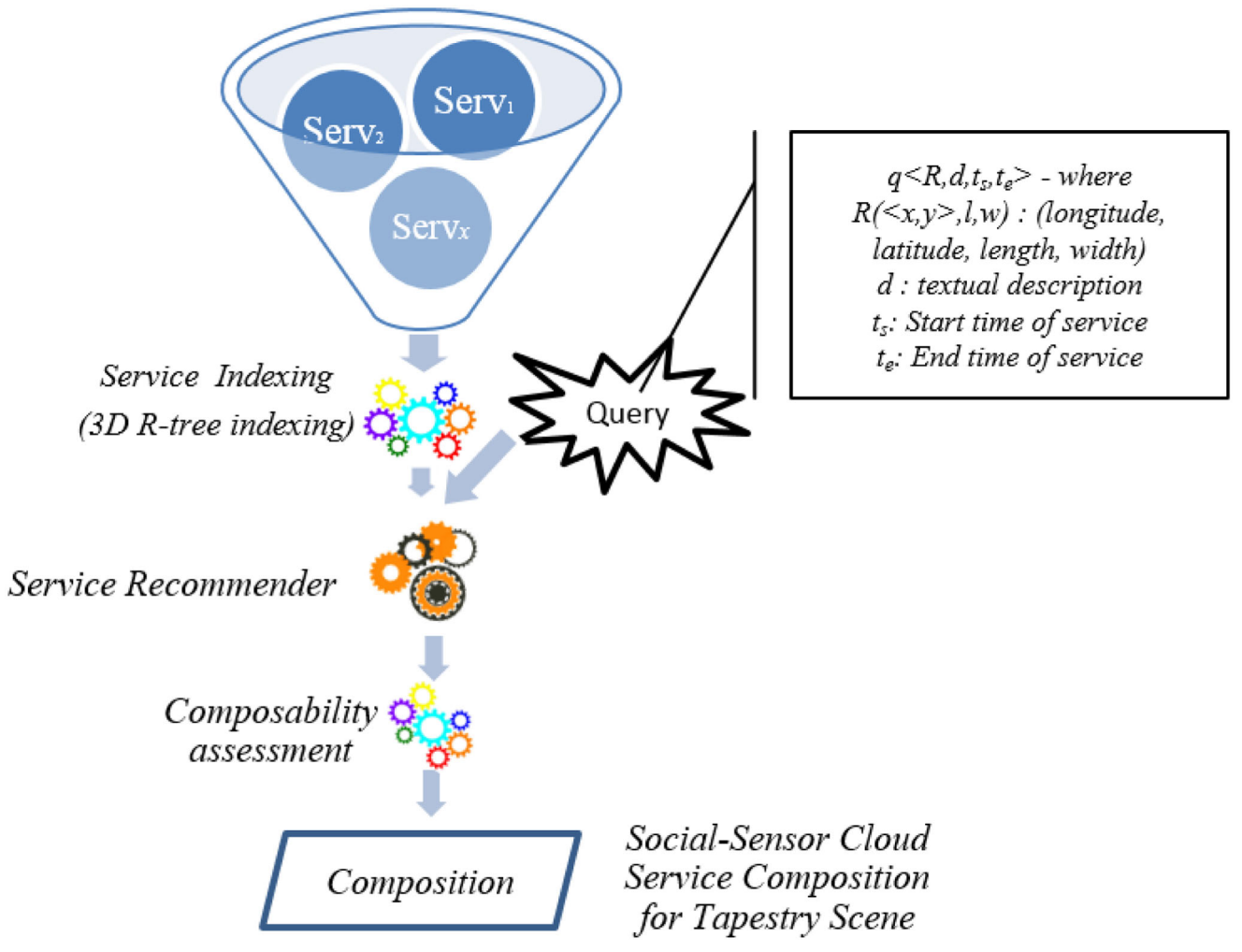

Fig. 3. Social-sensor cloud service composition process.

### 4.2 Functional Attributes of an Atomic Social-Sensor Cloud Service

The functional attributes capture the behaviours described by a service and form the baseline functionality of the *atomic service*. The following are the minimal functional attributes associated with an atomic service and their information sources:

- Social-sensor device
  - Shooting mode defines the nature of the service, such as square, video, panorama, etc.
- Social-sensor service owner
  - Description $D$ of the service provides additional information regarding the service. It is assumed that the service description includes complete details of the service specifics related to the scene captured, e.g., objects captured, and their relations.
- Social-sensor cloud
  - Tags $T$ are objects of interest in the service annotated by social media. It is assumed that the tags describe the details of the objects of interests and possible behaviours. Tags $T$ provide the location and focus of the image.

Description $D$ and tags $T$ together define the context *Con* of the sensor.

### 4.3 Non-Functional Attributes of an Atomic Social-Sensor Cloud Service

The non-functional attributes and sources of an atomic service include:

- Social-sensor device
  - Time $t$ is the time of the service at which the image is taken. $t$ can be either a single timestamp or a short interval of time $(t_s, t_e)$.
  - Location *L(x,y)* is the location of the service where $x$ is the longitude and $y$ is the latitude of the service.
  - Coverage *Cov* of the service defines the extent to which the service covers the scene. Coverage *Cov* is defined by VisD, dir, and $\alpha$, as shown in Fig. 4b. Where,

Visible distance *VisD* is the maximum coverage distance.
Direction *dir* is the orientation angle.
Angle $\alpha$ is the angular extent of the scene covered.

  - Resolution *Res* is the minimum requirement of image resolution to be provided by the service.
  - Colour quality *ColQ* is the image's definition, e.g., grey-scale.
- Social-sensor service owner
  - Accessibility *Acc* indicates if the service is publicly available to a broader audience.
  - Permanence *P* indicates if the service description is altered or not.
- Social-sensor cloud
  - Noise *N* is an obstruction in the image affecting the scene building as per user requirements. The SocSen service descriptions are used to detect noises. For example, 'pedestrians' can be considered as noises in coverage of vehicles on roads.

## 5 Social-Sensor Cloud Service Tapestry Composition Approach

We propose an approach to *filter*, *select* and *compose* the best available SocSen services to form an event summary for scene analysis. The composition aims to address the queries that are *ambiguous* regarding objects of interest and spatio-temporal attributes. The query $q$ is defined as q = $< A, t_s, t_e, d >$, giving the region of interest, description and quality parameters of the required service(s). A query sample is provided in Section 3. Our approach aims to *efficiently* compose the available services that match users' requirements by selecting *context relevant* and *composable* services using *functional and non-functional attributes*. The steps to manage and enable smooth SocSen service composition are depicted in Fig. 3. In the first step, we *index* all the available services using 3D R-tree [4] (Section 5.1). The second step is *service clusters recommendation* (Section 5.2). Then services are assessed for *composability* according to the predefined relations (Section 5.3). Finally, we *compose* a tapestry scene (Section 5.4).

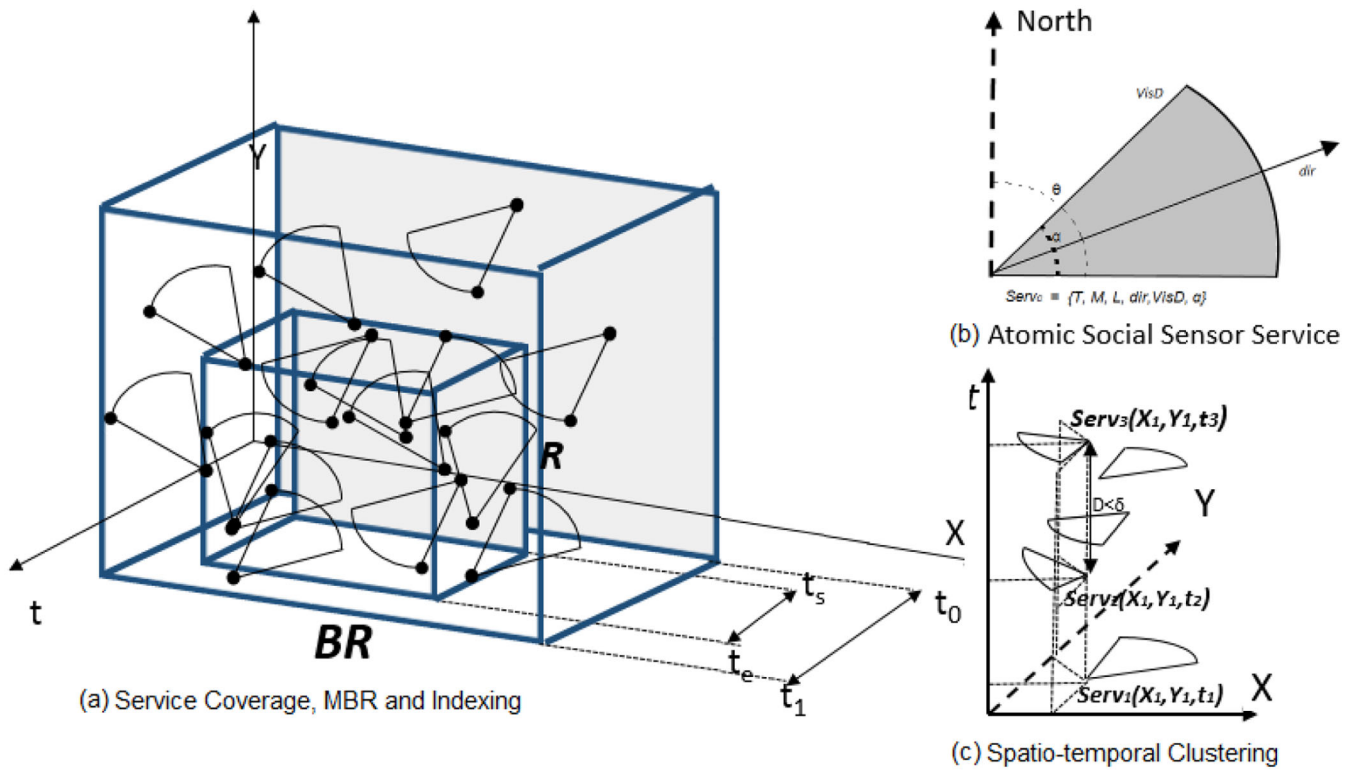

Fig. 4. Illustration of service coverage, *MBR*, indexing and spatio-temporal clustering.

There are two key assumptions. First, we assume that *all objects of interests and their behaviours are mentioned in the service description* (e.g., red car, erratic driving). Second, *none of the service coverage can wholly fit the contour of another*. The proposed technique forms a timeline of the queried event by constructing a tapestry of multiple smaller tapestries.

### 5.1 Social-Sensor Cloud Service Indexing

We need an efficient approach to select relevant images (i.e., services), given a large number of images in social media. Spatio-temporal indexing enables the fast discovery of services. We index services considering their spatio-temporal features using a *3D R-tree* [4]. 3D R-tree [6] is a tree data structure which can be used as an efficient spatio-temporal index to handle time and location-based queries. Time is considered as the third dimension in the 3D R-Tree. The leaf nodes of the 3D R-tree represent services which are organised using the minimum bounded region (MBR) that encloses the service spatio-temporal region (Fig. 4a). It is assumed that all available services are associated with a two-dimensional geo-tagged location and time. The 3D R-tree efficiently answers typical range queries for region $R < A, t_s, t_e >$, e.g., "select all services bounded by rectangular area $A$ in time $t_s$ to $t_e$".

### 5.2 Social-Sensor Cloud Service Recommender

The SocSen services are diverse in their coverage of space, time and content. Several problems are inherent to SocSen services due to such wide occurrence and free nature of social-sensors. Some of these problems include *lack of contextual similarity with required context, images only targeting sky or ground, services with much noise*. Such services are considered as irrelevant to the query. Extracting relevant services is crucial for composability assessment and composition of services. The purpose of the service recommender is to *select and recommend a set of clusters of services for the composition* by assessing spatio-temporal proximity and contextual relevance of the services. The contextually relevant spatio-temporal clusters help to analyse the spatio-temporal evolution of an event effectively. A key element in the design of a service recommender is *defining similarity metrics* to compare services. The service comparison strategy is based on the *meta-data* and *semantic annotations* of the images.

#### 5.2.1 Pre-Processing: Spatio-Temporal Clustering and Context Extraction

The recommendation of relevant services process entails 1) spatio-temporal clustering, and 2) context extraction for each cluster. We propose to cluster SocSen services from spatio-temporal and directional perspectives. A high-performance NLP API, *Retina*,[1] is used to extract the associated semantic context and keywords of each spatio-temporal cluster. Semantic similarities are based on the textual similarity between the semantic annotations of the service clusters and the query description. We explain the pre-processing steps as follows.

**Algorithm 1.** Spatio-Temporal Service Clustering

**Input:** 1) A set S of services - Each service has a spatio-temporal boundary and keywords. 2) Spatial distance threshold $\delta$. 3) Temporal distance threshold $\lambda$.
**Output:** A set $C_{ST}$ of spatio-temporally clustered services.

C = 1 ▷ Cluster Counter
**for** Each $Serv_i \in$ S **do**
  **Insert** $Serv_i$ in 3D R-Tree
**for** Each $Serv_i \in$ S **do**
  **if** $Title(Serv_i) \neq$ undefined **then** ▷ Check if a service is in any cluster or not
    $Title(Serv_i) = C$ ▷ Assign service to cluster
    $Serv_j \in C_{ST}.C$ ▷ $C_{ST}.C$ is a spatio-temporal cluster
    Form GeoPatch $\langle \alpha, p, \overrightarrow{dir}, VisD \rangle$ ▷ Form geographic spatial coverage with a direction and boundaries (Fig. 4b)
    S′ = Neighbours($Serv_i$, $\delta$, $\lambda$)
    **for** Each $Serv_j \in$ S′ **do**
      **if** $Direction(Serv_i, Serv_j)$ = TRUE **then** ▷ (Fig. 4c)
        $Title(Serv_j) = C$
        $Serv_j \in C_{ST}.C$
        S′ = S′ \ $Serv_j$
    C = C + 1 ▷ Next cluster label

*Spatio-Temporal Clustering*. Spatio-temporal clustering is a useful way to discover spatio-temporally and directionally relevant services efficiently. We propose to cluster SocSen services from spatio-temporal perspectives by considering the direction of the coverage. Clusters based on the vantage point of coverage help in handling multiple queries of similar context or coverage. Spatio-temporal clustering boosts the performance of the next step, i.e., composability assessment.

The following methodology is based on the coverage direction of the service and the distance between two services. The steps, as discussed in more details in Algorithm 1, are used to cluster SocSen services. The algorithm starts with an *arbitrary* service in the service set. The *neighbourhood services* of this service are extracted using a spatial distance $\delta$, temporal distance $\lambda$ and direction of the coverage (Fig. 4c). The directional relationships between the services depend upon the natural quadrants, i.e., North, South, East, and West. Two or more services are considered in a similar

1. http://www.cortical.io/API.html

direction when their directions from true north are considered in the same quadrant. If a service is in a new quadrant, the service becomes the first service in the new cluster. The algorithm then iteratively collects the neighbouring services within a given distance from the core service. The process is repeated until all the services have been processed. Neighbouring services as determined by euclidean distance

$$Dist_s(Serv_1, Serv_1) = \sqrt{(x_1 - x_2)^2 + (y_1 - y_2)^2} \quad (1)$$

$$Dist_t(Serv_1, Serv_1) = \sqrt{(t_1 - t_2)^2 + (t_1 - t_2)^2}. \quad (2)$$

*Context Extraction of Spatio-Temporal Clusters.* We extract a set of keywords $C_{ST.i}.kw$ from the cluster's services' descriptions and tags. It is assumed the social-sensor services' descriptions and tags provide information related to places, objects, buildings or behaviours captured in a service. Therefore, extracting context information from all the services available is essential. We *categorised* extracted keywords into four categories, i.e., objects of interest, landmarks, behaviours, and interacting objects. The service's owner provides service descriptions, including details of the service specifics related to the scene captured, e.g., objects captured, and their relations. Tags provided by social-sensor clouds are automatically generated and provide a *realistic* context. However, information from a SocSen service's owner is considerately *different* from tags provided by a social-sensor cloud. Information from a service's owner is mostly *subjective* and based on people's personal experience. We assume that data provided by the social-sensor cloud is *more trustworthy* than the description provided by the service's owner.

Retina API is used to extract keywords and annotations from service descriptions provided by service owners. Retina API parses the input text and numerically encodes the meaning of the text as a semantic fingerprint. Snippets are extracted based on their relation (i.e., semantic fingerprint) to the context of the service description. These snippets enable to extract keywords and annotations from text corpora by determining the importance of each term in the text. The context extraction process performs very accurate word sense disambiguation to find out what the user meant by each mentioned snippet. However, these snippets and their context are *subjective*. The Retina API lets us compare the meaning of terms on the semantic level. The semantic fingerprint representation also extends to representing the semantic content of texts in the same space using the terms as the basic unit. This gives the ability to compare the semantic content of two pieces of texts and extract the correct context as per service tags and coverage.

Service keywords *Serv.kw* are determined by analysing the overlap between the snippets from *Serv.D* and tags *Serv. T*. The analysis is based on semantic similarity between snippets and precedence of trustworthiness in a SocSen service. If a word exists in both tags and keywords, it is deemed as *Serv.kw*. The relationship between *Serv.kw*, tags and description snippets can be explained as

$$Serv.kw \in \begin{cases} Serv.T \\ Serv.D \;\; if \;\; Serv.D(snippet) \subset Serv.T \\ Serv.D \;\; if \;\; Serv.D(snippet) \cong Cluster.kw \end{cases} . \quad (3)$$

TABLE 2
Cluster Keywords and Similarity Scores

| Categories | Query Words | Cluster Keywords $C_{ST.1}$ | $C_{ST.2}$ | $C_{ST.3}$ | $C_{ST.4}$ |
|---|---|---|---|---|---|
| Object/s of Interest | Red Car | Red Car (1,0.4) | Red Car (1,0.6) | Red Car (1,0.2) | |
| Landmark | Road X, Road Y | Road X (1,0.5) Road Y (1,0.5) | Road X (1,0.5) | Road Y (1,0.2) | Road X (1,0.2) |
| Behaviour | Erratic Driving | Speeding (0.2,0.4) | Burnouts (0.3,0.8) Drove North (1,0.5) | Stuck Pedestrian (0,0.3) Mount Footpath (1,0.6) | Treating Pedestrian (0,0.2) |
| Interacting Objects | Emergency Service | | Pedestrian (N/A, 0.3) | | Ambulance (N/A, 0.5) |

We use Retina API for comparing the textual similarity between *Serv.d* and *Serv.T*. Retina parses the input text and numerically encodes the meaning of the two pieces of text as two semantic fingerprints. Further, it quantifies the meaning overlap between the two pieces of text as a percentage, based on the number of overlapping points between the items to compare. A snippet from *Serv.d* is identified as a service keyword if the contextual similarity, i.e., the overlap between the snippet and the cluster keywords is more than a set threshold $\theta$. *Serv.kw* represents a set of words extracted from the service tags and description that are likely to describe the service. We call the set of *kw* as event keywords.

$C_{ST.i}.kw$ are determined by the frequency of the unique *Serv.kw* in each cluster. Each $C_{ST.i}.kw$ is associated with its relative frequency in the cluster. The relative frequency of $C_{ST.i}.kw$ is calculated as

$$freq_{rel}(C_{ST.i}.kw) = \frac{freq(C_{ST.i}.kw)}{\sum_{i=0}^{n} freq(C_{ST.i}.kw)}. \quad (4)$$

We compute a keyword score indicating how likely a keyword matches the query words, as shown in Table 2. The score indicates how contextually similar the cluster keywords are with the query keywords. The overall context of the cluster can be assessed by analysing the distribution, i.e., context relevance and relative frequency of the service keywords in each spatio-temporal cluster. The context of this cluster can be used to depict what is happening in the area. Table 2 depicts the possible keyword categories as per the motivation scenario. Retina API for text comparison is used to assess semantic similarity.

#### 5.2.2 Spatio-Temporal and Contextual Cluster Recommendation

Contextual recommendation of the spatio-temporal service clusters around the location of the query is essential for the SocSen service composition. The purpose of the service recommender is to filter and recommend the spatio-temporally similar neighbours for the composition by assessing contextual relevance between services and queries. The significant elements of the proposed service recommender are as follow. 1) Assessment and filtering of the relevant services from the spatio-temporally and contextually relevant clusters. 2) Assessment of the spatio-temporal service clusters that are temporally close and follow similar semantic patterns. Algorithm 2 outlines the service recommendation process.

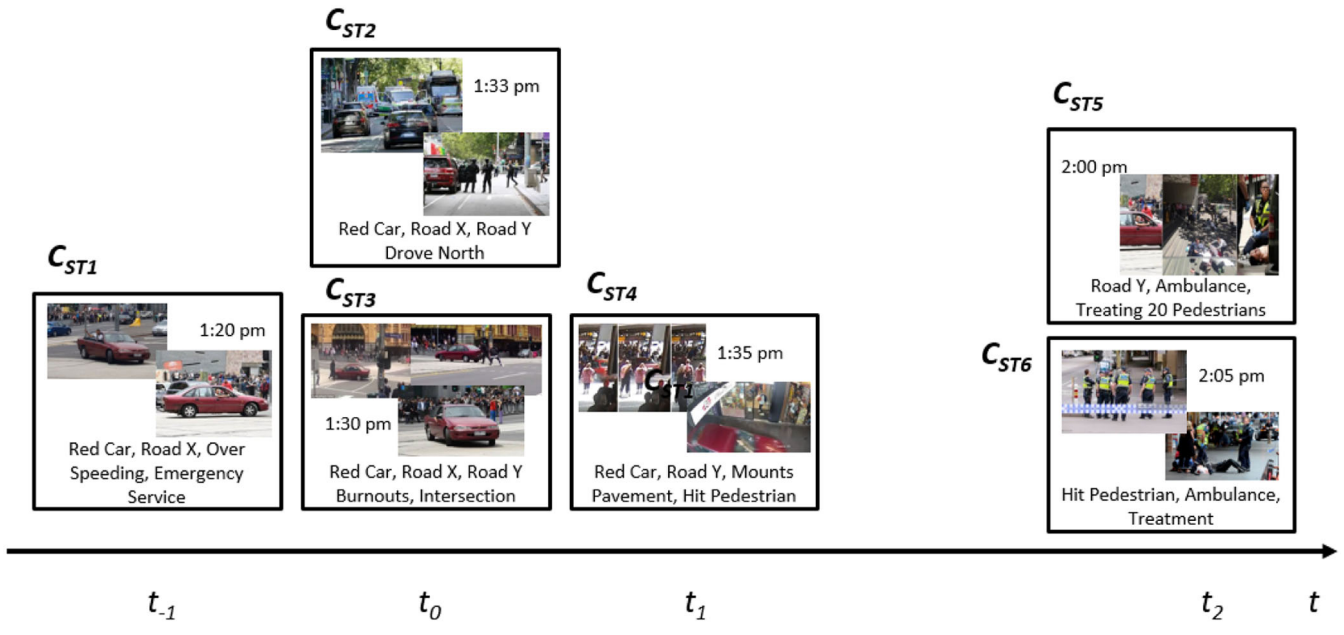

Fig. 5. Temporal-semantic similarity.

*Spatio-Temporal and Contextual Service Filtering.* Spatio-temporal and contextual filtering of the clusters relevant to the given query, within loose spatio-temporal bounds, is required for an effective recommendation. We define a cube shape region $R$ using the user-defined rectangular event area and the start and end time of the service, i.e., $t_s$ and $t_e$ for the effective area of the query $q$. The clusters outside this region are assumed to have little probability of being relevant to the query. Fig. 4a illustrates the query region and the bounded region $BR$ across time $\mathrm{t_s}$ to $\mathrm{t_e}$ forming MBR. The bounded region $BR$ is formed by rounding up the query region. Next, the semantic similarity between the clusters and a query is assessed. Top $n$ keywords of each cluster, $C_{St}.kw_0^n$, according to the relative frequency are used to assess the semantic similarity between services in the clusters and the query. LIN's semantic similarity model [21] is used to calculate the textual similarity between $C_{St}.kw_0^n$ and a $q.d$. A cluster is identified as contextually relevant to the query if the similarity between a $q.d$ and cluster keywords is more than a set threshold $\theta$. Next, the contextually irrelevant services from each spatio-temporally and contextually relevant cluster are filtered out. Again, LIN's model [21] is used to assess contextually irrelevant services. $\theta$ is used again as a similarity threshold to assess the contextually irrelevant services.

*Temporal-Semantic Similarity.* Temporally and semantically similar spatio-temporal services clusters are used to represent an event's *activity pattern* (Fig. 5). This activity pattern depicts the *spatio-temporal evolution* of the event. This approach leverages temporal information as a complementary source to detect semantic relatedness and computes a refined metric of semantic relatedness. This approach is based on associating each cluster keyword set with a *weighted frequency of similarity*. The weighted frequency of similarity $S_{sim}$ is based on their unified occurrence and calculated as

$$S_{sim}(C_{ST}.x, C_{ST}.y) = \left| \frac{freq(C_{ST}.x[kw] \cap C_{ST}.y[kw])}{freq(C_{ST}.x[kw] + C_{ST}.y[kw])} \right|. \tag{5}$$

Next, the *temporal dynamics* for each context is extracted to quantify context occurrence for each period (e.g., within 5 min of the desired time-stamp) to build its time series. Finally, each context's time series is *scaled* according to the context's original weight from Retina engine.

*Welch's t-test* is applied to compute the similarity in terms of time. Welch's t-test dynamically assesses time associated with two clusters for similarity measure based on the characteristics of services. Welch's t-test is defined as

$$T_{sim}(C_{ST}.x, C_{ST}.y) = \frac{\overline{\overline{x_1} - \overline{x_2}}}{\sqrt{\frac{s_1^2}{N_1} + \frac{s_2^2}{N_2}}}. \tag{6}$$

Where, $\overline{x_1}$, $s_1^2$ and $N_1$, and $\overline{x_2}$, $s_2^2$ and $N_2$ are sample mean, sample variance and sample size of two clusters, respectively. Welch's t-test is applied to compare activity distributions in terms of time for every two clusters.

The temporal-semantic relation between clusters is

$$Relation(C_{ST}.x, C_{ST}.y) = \begin{cases} \frac{S_{sim}(C_{ST}.x, C_{ST}.y)}{|T_{sim}(C_{ST}.x, C_{ST}.y)|} \\ 0 \;\; if \;\; \lim_{S_{sim}\to 0} \\ 0 \;\; if \;\; \lim_{T_{sim}\to \infty} \end{cases}. \tag{7}$$

Where, 1 is considered the maximum temporal-semantic similarity, and 0 is considered as no temporal-semantic similarity. For example, $T_{sim}(C_{ST}.2, C_{ST}.3) = \frac{\overline{\overline{33-30}}}{\sqrt{\frac{1}{2}+\frac{1}{3}}} = 3.61$, as calculated through Welch's t-test. The keyword overlap for $C_{ST}.2$ and $C_{ST}.3$ is 3 out of 5 keywords. *Relation* $(C_{ST}.2, C_{ST}.3) = \frac{3}{|3.61|} = 0.83$. Therefore, it can be stated that they have a high temporal-semantic similarity. The temporal-semantic similarity is lower for the clusters $C_{ST}.1$ and $C_{ST}.2$. $T_{sim}(C_{ST}.1, C_{ST}.2) = \frac{\overline{\overline{20-33}}}{\sqrt{\frac{1}{2}+\frac{1}{2}}} = 13$. The keyword overlap is 2 out of 5 keywords. $Relation(C_{ST}.1, C_{ST}.2) = \frac{2}{|13|} = 0.15$.

**Algorithm 2.** Service Recommendation

**Input:** 1) A set S of $C_{ST}$ - Spatio-temporally clustered services. 2) A query $q < R(A, t_s, t_e), d >$. 3) A time period $(t_1, t_2)$ for the maximum search of related events.

**Output:** 1) A set $S'$ of $C_{STC}$ - Spatio-temporally and contextually relevant clusters. 2) A set $S^+$ of $C_{STC}$ - Temporal and contextual relevant clusters.

1: **for** Each $C_{ST} \in$ S **do** Select spatio-temporally and contextually relevant clusters
2: **if** ($C_{ST} \cap$ R) $\wedge$ (($C_{ST}.kw_0^n \sim q.d) \geq \theta$) **then**
3: $C_{ST} \in S'$
4: **for** Each $C_{ST} \in$ S′ **do** Filter out contextually irrelevant services from clusters
5: **for** Each ($Serv_i \in C_{ST.i}$) **do**
6: **if** ($Serv_i.kw \sim q.d) \leq \theta$ **then**
7: $C_{ST.i} = C_{ST.i} \setminus Serv_i$
8: $C_{ST.i}.kw = C_{ST.i}.kw \setminus Serv_i.kw$
9: **for** Each $C_{STC.i} \in$ S′ **do** Extract temporal-semantic similar clusters
10: S″ = Neighbours($C_{STC.i}$, $(t_1, t_2)$)
11: **for** Each $C_{STC.j} \in$ S″ **do**
12: **if** $C_{STC.j}.kw_0^n \sim C_{STC.i}.kw) \geq \theta$ **then**
13: $C_{STC.j} \in S^+$ Temporal-semantic similar clusters

### 5.3 Social-Sensor Cloud Service Composability

We propose a spatio-temporal composability model for SocSen services. Spatio-temporal and contextual composability of two or more SocSen services can be defined in terms of three instances:

- Two or more services are composable if these services are spatio-temporally composable and contextually relevant.
- Two or more services are composable if these services are temporally composable and contextually relevant. In such cases, services might be located outside the region of interest but still capture a scene inside.
- Two or more services are composable if these services are spatially composable and contextually relevant. In such cases, services are available either before or after the required period. Therefore, these services might not directly capture the event in the defined period but help in building the scene for the cause and effect of the event.

#### 5.3.1 *Temporal Composability*

Temporal composability, i.e., $Comp_t$, of two or more SocSen services, defines the temporal relationship between the services. These temporal relationships are defined by services' time stamp $T$ or time interval defined by start time $T_s$ and end time $T_e$. The time $T$ is the timestamp of the service when a stand-alone image is taken. Start time $T_s$ and end time $T_e$ define the interval $(T_s, T_e)$ of the service for short video format pictures (e.g., gif). Based on the above description, the following rules for temporal composability are defined.

Let $Serv_1$ and $Serv_2$ be two services, then the expression $A$ $oper$ $B$ represents all the temporal relationships between the two services, where $oper \in \{>, <, =\}$. $Serv_1$ and $Serv_2$ are composable if:

- $Serv_1$ and $Serv_2$ touch temporal boundaries. This encompasses: 1) meet, i.e., $Serv_1.T_e = Serv_2.T_s$, 2) start together, i.e., $Serv_1.T_s = Serv_2.T_s$ and 3) end together, i.e., $Serv_1.T_e = Serv_2.T_e$,
- $Serv_1$ fully overlaps $Serv_2$, e.g., $(Serv_1.T_s > Serv_2.T_s)$ AND $(Serv_1.T_e < Serv_2.T_e)$,
- $Serv_1$ partially overlaps $Serv_2$, e.g., $(Serv_1.T_s < Serv_2.T_s)$ AND $(Serv_1.T_e < Serv_2.T_e)$ AND $(Serv_1.T_e > Serv_2.T_s)$,
- $Serv_1$ and $Serv_2$ are equal, e.g., $(Serv_1.T_s = Serv_2.T_s)$ AND $(Serv_1.T_e = Serv_2.T_e)$., or
- $Serv_1$ and $Serv_2$ have the minimum temporal distance between them. This encompasses $Serv_1.T_e = Serv_2.T_s + \Lambda$, where $\Lambda$ is the maximum allowed difference in time.

Euclidean distance is used as a distance function for the time difference between the two services (Equation (2)).

#### 5.3.2 *Spatial Composability*

Another aspect of composition concerns spatial relationships between two or more services. The spatial composability of two or more services $Comp_s$ is defined by:

- The directional relationships between the services, e.g., North, South and South-East, etc. Two or more services are considered topologically composable when their directions from true North are considered in the same quadrant of the 4-quadrant general directional model, and
- The distance between the services, e.g., the shortest distance between the two services.

Let $Serv_1$ and $Serv_2$ be two service. $Serv_1$ and $Serv_2$ are composable if:

- $Serv_1$ and $Serv_2$ touch spatial boundaries. This encompasses: 1) Overlap in space, or 2) Have the minimum spatial distance between them, i.e., close in geo-locations. This encompasses $Serv_1.L = Serv_2.L \pm \Delta$, where $\Delta$ is the maximum allowed difference in geo-locations. euclidean distance is used as the distance function for the two services $Serv_1$ and $Serv_2$ (Equation (1)). And
- $Serv_1$ and $Serv_2$ share the vantage point, i.e., $Serv_1$ and $Serv_2$ lie in the same geometric quadrant, and are spatially close. This encompasses $(Serv_1.Cov_{(\alpha,dir)} \cong Serv_2.Cov_{(\alpha,dir)})$, i.e., similar in coverage.

The pictorial representation of the spatial composability relation is shown in Table 3.

TABLE 3
Spatial Composability

| ***Spatial Relationship*** | ***Pictorial Representation*** | ***Operator Relationship*** |
|---|---|---|
| Topological and Directional Relationship | 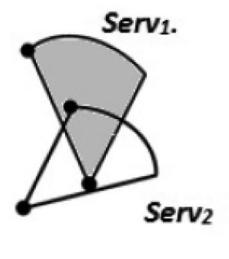  | $Serv_1.Cov_{(\alpha,dir)} \cap Serv_2.Cov_{(\alpha,dir)}$<br>AND<br>$0^\circ < Serv_1.dir < 90^\circ$ **AND** $0^\circ < Serv_2.dir < 90^\circ$<br>***OR***<br>$90^\circ < Serv_1.dir < 180^\circ$ **AND** $90^\circ < Serv_2.dir < 180^\circ$<br>***OR***<br>$180^\circ < Serv_1.dir < 270^\circ$ **AND** $180^\circ < Serv_2.dir < 270^\circ$<br>***OR***<br>$270^\circ < Serv_1.dir < 360^\circ$ **AND** $270^\circ < Serv_2.dir < 360^\circ$ |
| Distance Relationship | 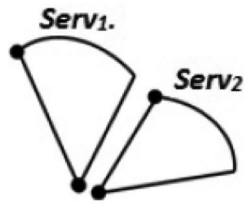  | $Serv_1.L = Serv_2.L \pm \Delta$<br>*Where Δ is the maximum allowed distance between geo-locations.* |

#### 5.3.3 *Composability Assessment*

All services within a cluster are considered composable because they share the same *vantage point* and *minimum spatial distance*. In addition, two clusters are considered composable if *any two services from both the clusters are composable*. Therefore, *spatial and temporal composability* of the services from the different clusters is required to assess the composability relations between the clusters. Temporal composability (i.e., $Composability_T$) between two services is the temporal overlap (i.e., $Overlap_T$) of the services. It is assessed as

$$Overlap_T(Serv_1, Serv_2) = \frac{Dist_t(Serv_1, Serv_1)}{\Lambda}. \tag{8}$$

Where, $Overlap_T(Serv_1, Serv_2) = 1$ is considered as the maximum acceptable temporal overlap for the temporal composability.

Spatial composability (i.e., $Composability_S$) between the services is a qualitative relationship. We assess the composability using the spatial overlap (i.e., $Overlap_S$), positioning distance and direction of the services. Spatial overlap helps to assess the coverage overlap of the service

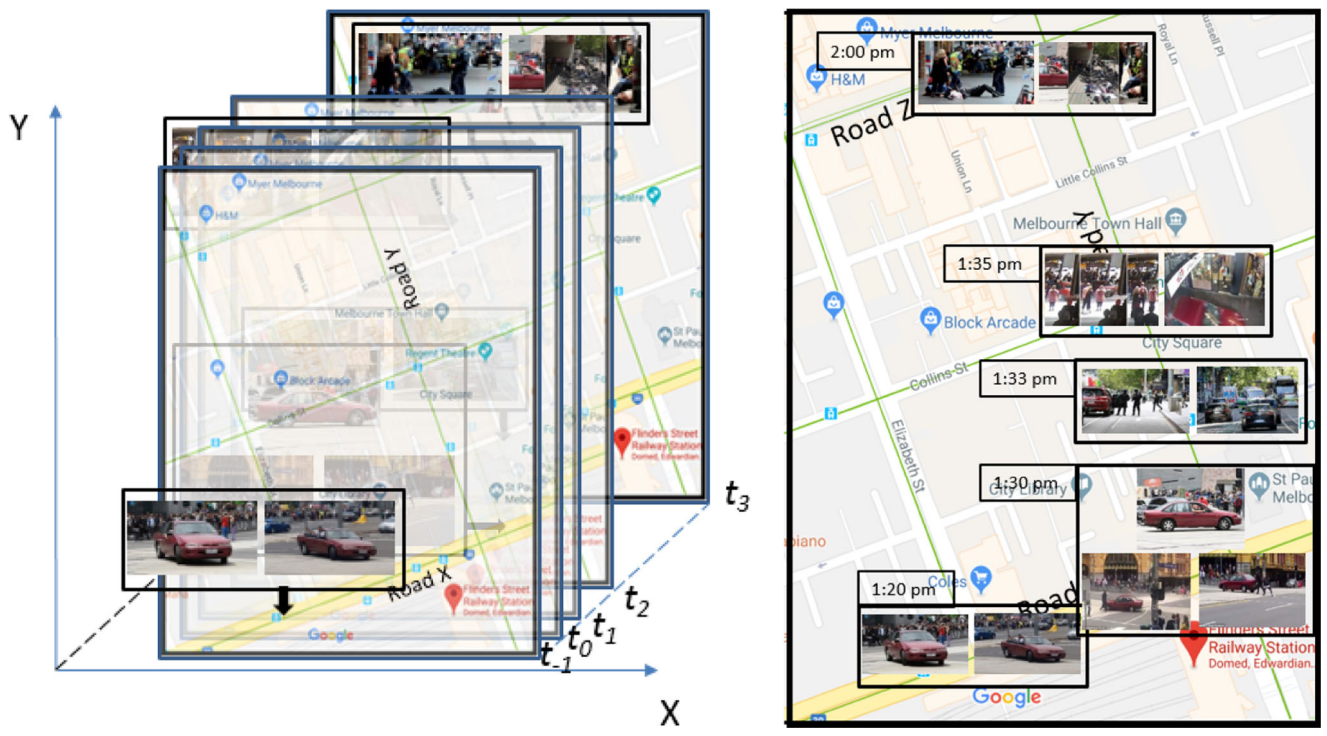

(a) Portrayal in 3D (b) Portrayal in 2D

Fig. 6. Service composition portrayals.

$$Overlap_S(Serv_1, Serv_2) = \frac{1}{2}\,|\,(Serv_1.VisD^2 * Serv_1.\alpha - Serv_2.VisD^2 * Serv_2.\alpha\,|\,. \quad (9)$$

Where, $Overlap_S(Serv_1, Serv_2) = 0$ depicts the maximum coverage area overlap

$$Overlap_{\overrightarrow{dir}}(Serv_1, Serv_2) = |\,Serv_1.\overrightarrow{dir} - Serv_2.\overrightarrow{dir}\,|\,. \quad (10)$$

Where, $Serv_1.\overrightarrow{dir}$ and $Serv_2.\overrightarrow{dir}$ are direction angles of $Serv_1$ and $Serv_2$ respectively. The maximum acceptable distance overlap for composability is $\Delta$. $\vartheta$ is the maximum acceptable direction overlap for $\overrightarrow{dir}$. Therefore, services are considered spatio-temporally overlapped if the difference between distance and direction of both services is below the threshold. $Composability_S(Serv_1, Serv_2)$ is depicted as

$$Composability_S(Serv_1, Serv_2) \begin{cases} = 1 \ \ if\ 0 < \frac{Dist_t(Serv_1, Serv_1)}{\Delta} \leq 1 \wedge \\ Overlap_{\overrightarrow{dir}}(Serv_1, Serv_2) < \vartheta^o \\ = 0 \ \ if\ Overlap_S(Serv_1, Serv_2) = 0 \wedge . \\ Dist_t(Serv_1, Serv_1) = 0 \wedge \\ Overlap_{\overrightarrow{dir}}(Serv_1, Serv_2) = 0^o \end{cases} \quad (11)$$

Spatio-temporal composability, $Composability_{ST}(Serv_1, Serv_2)$, is based on the time of the services, their proximity in space and the contextual similarity between service keywords (i.e., $Serv_1.kw \sim Serv_1.kw$). Composability is assessed on the basis of relations described at the beginning of the section. These relations are depicted as

$$Composability_{ST}(Serv_1, Serv_2) = 1 \begin{cases} if\ Composability_S = 1 \\ \ \ \wedge\ Composability_T = 1 \\ if\ Composability_S = 1 \\ \ \ \wedge\ (Serv_1.kw \sim Serv_1.kw) > \theta \\ if\ Composability_T = 1 \\ \ \ \wedge\ (Serv_1.kw \sim Serv_1.kw) > \theta \end{cases} . \quad (12)$$

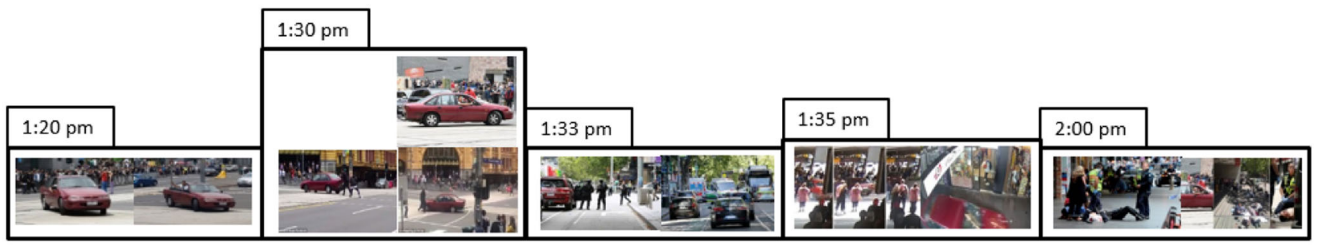

Fig. 7. Event timeline portrayal of a tapestry scene.

## 5.4 Social-Sensor Cloud Service Composition for Tapestry Scenes

The composition is handled as *sewing a tapestry* to form the scene. We start with the *central* piece, concerning space and time, and build a tapestry around it. The build-up is based on selecting the most composable services from the set of recommended services. The tapestry scene covers the visual summary of the whole queried *scene*, i.e., all objects of interest and their context. The final composition ascertains the uncertain parameters like *all events relevant to the user's query* and *previously undiscovered objects of interests involved*. Moreover, keywords describing smaller events, objects of interests and related contexts are also generated.

The proposed solution forms a tapestry of multiple scenes, forming a timeline of the queried event. The services' coverage does *not* fit wholly with the contour of each other because of the dissimilar and divergent nature of the services' coverage. In the proposed composition approach (Algorithm 3), we first spatially *grid* the query area into four natural cartesian plane quadrants using the query location as a central reference. Next, the recommended cluster that is closest to the query area is *picked* and *placed* in the gridded query area. Further, neighbour clusters are *picked*, and composable services are *patched* together according to their proximity in the cartesian plane. Fig. 6a depicts a 3D portrayal of the SocSen service composition. Fig. 6b depicts the SocSen service composition in a 2D plane along the queried period. The process of selecting and joining the services continues until all composable services are patched to provide the maximum available coverage. This process forms a 2D portrayal of the required composite service as a tapestry scene. Spatial and temporal gaps in the final output are expected, depending upon the services available in the queried area and time interval. The final 2D rendition of the composite service is a series of spatial tapestries ordered by time, providing a timeline of the visual summary of the event. Fig. 7 depicts the visualisation of the finalised tapestry scene.

**Algorithm 3.** Service Composition for Tapestry Scenes

**Input:** A set $S^*$ of $C_{STC}$ - Spatio-temporal and contextual composable clusters.

**Output:** 1) A set $S^{**}$ of $C_{comp}$ - Service composition portrayal in 2D. 2) A composed service $Serv_{comp}$ - Event timeline portrayal of the tapestry scene.

1: **for** Each $C_{STC.i} \in S^*$ **do**
2:   **for** Each $Serv_j \in C_{STC.i}$ **do**
3:     Stitch GeoPatch($Serv_j$) Stitch $Serv_j$ to GeoPatch (Algorithm 1 Step 8) by joining spatially to the patch
4:   $S^{**} = S^{**} \cup C_{comp.i}$ $S^{**}$ is a set of clusters composed as a service
5: $Serv_{comp} = Sort_t(S^{**})$ Sort all clusters in a temporal order

TABLE 4
Relevance @ $\delta$ for Query 1

| $\delta$ | Cluster 1 | Cluster2 | Cluster3 | Average |
|---|---|---|---|---|
| 20 | 0.3750 | 0.8333 | 0.44441 | 0.5509 |
| 40 | 0.3333 | 0.3684 | 0.2222 | 0.3080 |
| 60 | 0.4286 | 0.2581 | 0.0800 | 0.2555 |
| 70 | 0.2182 | 0.0127 | 0.0114 | 0.0807 |
| 80 | 0.1600 | 0.0000 | 0.0000 | 0.0533 |
| 90 | 0.1364 | 0.0000 | 0.0102 | 0.0489 |

TABLE 5
Relevance @ $\delta$ for Query 7

| $\delta$ | Cluster 1 | Cluster 2 | Cluster 3 | Average |
|---|---|---|---|---|
| 20 | 0.3333 | 0.4000 | 0.5556 | 0.4296 |
| 40 | 0.3913 | 0.9091 | 0.1111 | 0.4705 |
| 60 | 0.4839 | 0.5357 | 0.3200 | 0.4465 |
| 70 | 0.6000 | 0.5536 | 0.7624 | 0.6386 |
| 80 | 0.6515 | 0.7500 | 0.8061 | 0.7359 |
| 90 | 0.4944 | 0.5000 | 0.7025 | 0.5656 |

# 6 Experiment and Evaluation

A set of experiments is conducted to evaluate and analyse the contribution of our proposed approach.

## 6.1 Experimental Setup

To the best of our knowledge, there is no available real spatio-temporal service test case to evaluate our approach. Therefore, we focus on evaluating the proposed approach using a self-collected real dataset. The dataset is a collection of 10,000 user-uploaded images from January 2016 to July 2019, in the city of Melbourne, Australia. The images are downloaded from multiple social media websites, e.g., flicker, twitter, google+. We retrieve and download the images related to different keywords, in different locations and time windows. The query words can be found in the "Event Theme" column in Table 7. These retrieved images are spatially overlapped in nature. We use their geo-tagged locations, the time when an image was captured, posted descriptions and tags to create the SocSen services for the images. Further, camera direction $\overrightarrow{dir}$, maximum visible distance of the image *VisD* and viewable angle $\alpha$ are used to model the non-functional attribute *Cov*. Quality features, e.g., resolution, are abstracted as the other non-functional attributes. The user of the service can adjust the thresholds of the QoS parameters.

All the experiments are implemented in Java and Matlab. All the experiments are conducted on a Windows 7 desktop with a 2.40 GHz Intel Core i5 processor and 8 GB RAM. We conduct the experiment based on seven different *event-oriented queries based on the real events* in our dataset. The basic features of the queries are briefly introduced in Table 7. The example of a detailed user query can be found in Section 3. We have performed two sets of experiments, respectively, for configuration and evaluation. First, we tune the parameters of the proposed approach to obtain better performance (Section 6.2). Next, we evaluate the proposed approach by comparing its performance with several existing approaches from the perspectives of effectiveness and scalability (Section 6.3).

## 6.2 Parameter Tuning

We need to properly configure the parameters in the algorithms for better performance of the proposed approach. The process of parameter configuration entails:

### 6.2.1 Configuration of the Cluster's Spatial Distance $\delta$, and the Temporal Distance $\lambda$

Appropriate configurations of spatial distance $\delta$ and temporal distance $\lambda$ help to run the service clustering algorithm efficiently. If $\delta$ is low, clusters might not be able to capture relevant services. If $\delta$ is high, clusters may contain many irrelevant services. Similarly, an inappropriate configuration of the time period $\lambda$ results in clustering of services that are not contextually relevant but share the same spatial bounds. For example, two separate accidents occurred at the same road intersection are clustered together if $\lambda$ is too high. Therefore, a trade-off is needed.

We assess the effect of different values of $\delta$ on the relevance of services within service clusters for all the seven queries. As an example, we present the results of two different queries. The first query has an initial area of $100\ \mathrm{m}^2$ in the city centre ( Table 7 - Query 1). The second query has an initial area of $250\ \mathrm{m}^2$ in the suburban area (Table 7 - Query 7).

We test 6 different values of $\delta$. We select three clusters closest to the central point of the query for each $\delta$. We assess the relevance of the services in each cluster to the query. The results depict that $\delta$ impacts differently for the different types of spatial areas (Tables 4 and 5). In the first query (Table 4), $\delta$ = 20 gives the highest relevance. That is, the services within 20 meter are considered to be in the same clusters for the congested city centre area. The most appropriate distance $\delta$ is found to be 80 meters for the second query because the event density is low in the suburban areas (Table 5). Therefore, it is concluded that $\delta$ varies upon the service and event density in the spatial area. Similar to $\delta$, $\lambda$ can be varied based on the timespan of available images and videos. However, we do not have opportunities to testify the impact of $\lambda$ as we only use *images* in this experiment. $\lambda$ is set to a single timestamp. The clusters are created based on each single timestamp.

### 6.2.2 Configuration of the Composability Thresholds $\Delta$, $\vartheta$ and $\Lambda$ for Service Composition

It is essential to properly select the composability thresholds enabling higher accuracy of the composition. Appropriate configurations of spatial distance threshold $\Delta$, direction angle threshold $\vartheta$ and temporal distance $\Lambda$ assist to select spatio-temporally and contextually composable clusters efficiently. All the services within a cluster are considered *composable* since they share the same vantage point and the minimum spatial distance (Section 5.3.3). Therefore, we use each cluster's appropriate spatial distance $\delta$ (obtained in Section 6.2.1) as $Composability_S$'s appropriate distance threshold $\Delta$ to maintain the uniform composability rules. We ignore the impact of $\vartheta$ as it purely relies on *sparsity* of images. Properly configuring $\Lambda$ results in composing the services that are not close to each other in space but might share the same context. For instance, two accidents that occur in different locations and close time periods might be related to each other in context. Therefore, the tempo-contextual relation between spatio-temporally and contextually relevant clusters is used as composability defining metric.

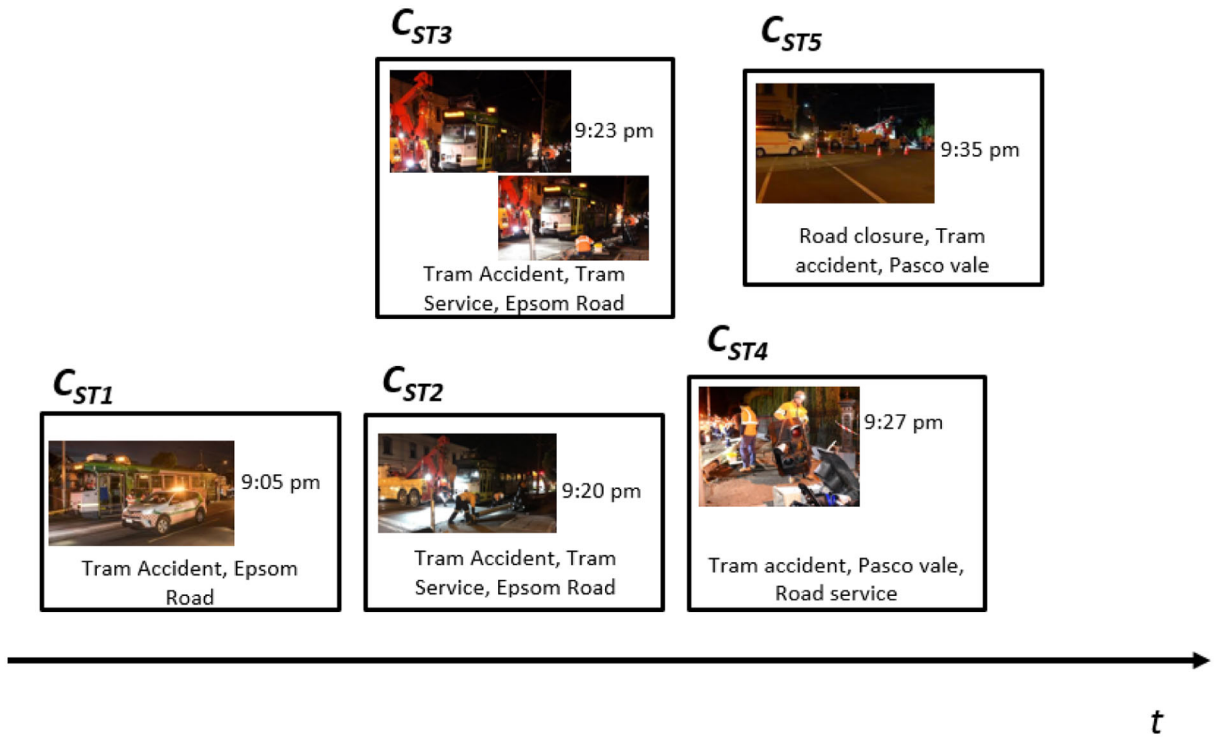

Fig. 8. Event timeline for Query 2.

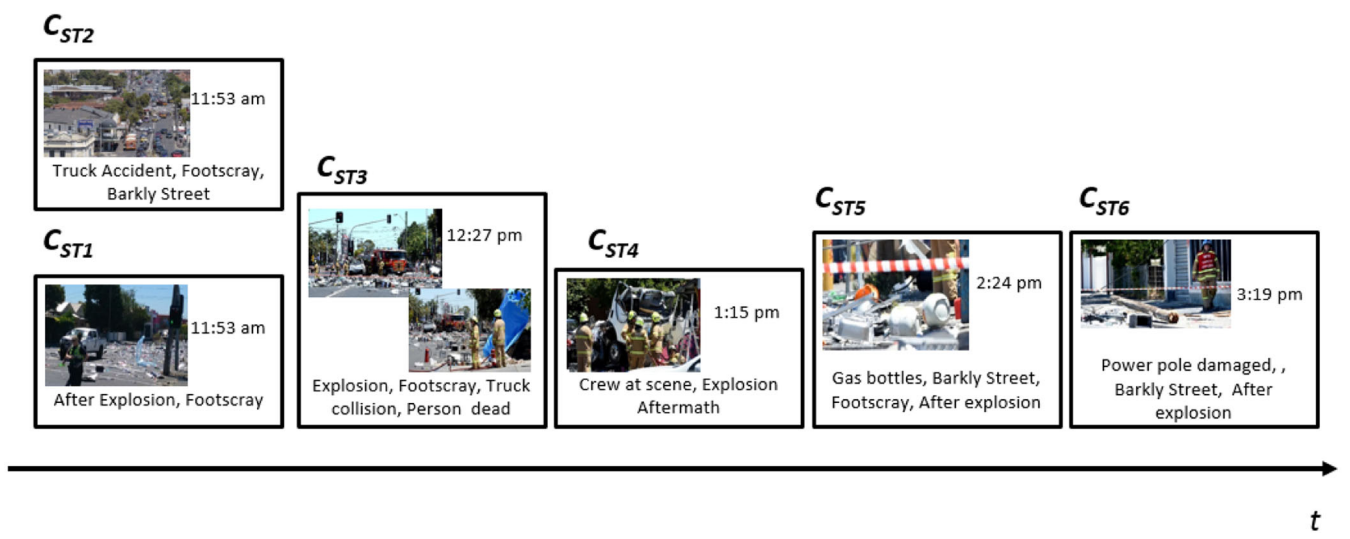

Fig. 9. Event timeline for Query 3.

We analyse the appropriate thresholds for $\Lambda$ to get better event coverage for the queries. We analyse two different types of queries, i.e., *isolated events* versus *chained events*. We assess the effect of different values of $\Lambda$ on composing the clusters into a single service for each query. As an example, we present the results of three different queries. The first query is about a single accident (i.e., overspeeding of a red car at Road X). This accident has temporal-semantic relation with another isolated accident, i.e., the accident of a red car on Road Y (Fig. 5). The second query is an isolated event, capturing a tram accident (Fig. 8). The third query is also an isolated accident (i.e., a gas explosion). However, the user of the query intends to capture the complete after-effects of the accident (Fig. 9).

We test 6 different values of $\Lambda$ for each query. We assess the number of the relevant clusters in each time period. The result (Table 6) indicates that $\Lambda$ relies on the query types. For example, services within close time proximity are enough for completing the visual summary of the scene, when composing for an isolated event. The experimental results (Query 2 and 3) show that 15 minutes time period can provide an appropriate visual summary of the tapestry scene for isolated events. The number of the clusters relevant to the queried isolated events remains constant for $\Lambda > 15$. Therefore, it is concluded that $\Lambda$ is set as 15 minutes for isolated events. However, a different $\Lambda$ is required for an event made up of several smaller events in different spatial locations, e.g., Query 1. For chained events, it would be better set $\Lambda$ either by domain experts as per the queried domain or by users of queries. In addition, $\Lambda$ might need adjustments depending on the context needed to be covered as per a user query (e.g., Query 3). Table 7 depicts the appropriate composability thresholds for each query in our experiments.

TABLE 6
Relevant Clusters @ $\Lambda$

| $\Lambda$ | Query :1 | Query 2 | Query 3 |
|---|---|---|---|
| 5 | 2 | 3 | 2 |
| 10 | 3 | 4 | 2 |
| 15 | 5 | 4 | 2 |
| 20 | 5 | 4 | 2 |
| 25 | 5 | 4 | 2 |
| 30 | 5 | 5 | 3 |

TABLE 7
Configured Parameters for the Queries

| | Event Type | Event Theme | Query Context | Query's Bounded Region (BR) | Δ | Λ |
|---|---|---|---|---|---|---|
| Q1 | Chained | Bourke Street Car Accident | Cause and Effects | $100m^2$ | 20m | 30mins |
| Q2 | Isolated | Pascovale Tram Derail | Accident Coverage | $100m^2$ | 20m | 15mins |
| Q3 | Isolated | Footscray Truck Blast | Cause and Effects | $500m^2$ | 30m | 30mins |
| Q4 | Isolated | Elizabeth Street Car Accident | Accident Coverage | $100m^2$ | 80m | 15mins |
| Q5 | Chained | Flinder Street Car Burnouts | Accident Coverage | $150m^2$ | 20m | 15mins |
| Q6 | Isolated | Essendon Airport Crash | Accident Coverage | $1000m^2$ | 70m | 15mins |
| Q7 | Isolated | Footscray Bridge Bus Crash | Accident Coverage | $250m^2$ | 20m | 15mins |

TABLE 8
Configuration of the Semantic Relatedness Threshold $\theta$

| $\theta$ | Accuracy | Relevance | Harmonic Mean |
|---|---|---|---|
| 0.25 | 80% | 65% | 72% |
| 0.50 | 69% | 77% | 73% |
| 0.75 | 55% | 77% | 64% |
| 1.00 | 51% | 88% | 64% |

#### 6.2.3 Configuration of the Semantic Relatedness Threshold $\theta$ for Service Selection

We aim to configure the appropriate value of $\theta$ for service selection. The evaluation is conducted in terms of *accuracy* and *relevance* of selected services for different semantic similarity thresholds. We conduct the experiment with different contextual semantic similarity thresholds $\theta$ (Section 5.2.1), (i.e., 0.25, 0.5, 0.75 and 1.0). The result (Table 8) shows that if the similarity threshold decreases, the number of the relevant services that are successfully selected (i.e., accuracy) increases, but the number of selected services that are relevant to the query (i.e., relevance) decreases. In contrast, the relevance increases if the similarity threshold is increased. However, the accuracy decreases simultaneously. Therefore, the similarity threshold $\theta$ is set to 0.5 based on the highest harmonic mean.

### 6.3 Evaluation

The experiments aim to evaluate the proposed approach from three perspectives. First, we evaluate the *accuracy* and *relevance* in spatio-temporal and contextual coverage according to the query requirements. Second, we analyse and evaluate the *scalability* of the proposed approach over time and space. Third, we conduct an online survey to find out human perception towards the output quality of the proposed approach.

TABLE 9
Composition Performance Comparison

| Approach | Accuracy | Relevance | Harmonic Mean |
|---|---|---|---|
| Proposed Approach | 69% | 80% | 74% |
| GCV | 28% | 79% | 42% |
| SIFT | 29% | 39% | 34% |
| KAZE | 35% | 56% | 43% |
| Selection Approach | 65% | 73% | 69% |

### 6.3.1 Accuracy and Relevance of Service Composition

The first set of experiments assesses the accuracy and relevance of composition. The accuracy of composition is assessed by *the ratio of the successfully retrieved services to the services relevant to the query*. The relevance of the composition is calculated as *a fraction of the successfully retrieved relevant services to the services relevant to the query*. The harmonic mean of accuracy and relevance is used to assesses the best available solution.

We compare the proposed composition approach with four existing approaches for evaluation. The existing approaches include 1) classic image processing technique of Scale-Invariant Feature Transform [16] (abbreviated as *SIFT*), 2) enhanced image processing technique using KAZE features [31] (abbreviated as *KAZE*), 3) Image analytic based on Google Cloud Vision API[2] (abbreviated as *GCV*) and 4) The selection based composition approach [5] (abbreviated as *the selection approach*). SIFT and KAZE are both implemented by similarity analysis. Their major differences lie in targeted features and matching algorithms. We use a 360 degree structured image dataset *I* from Google Map Street View of the area of interest for each query. The similarity analysis is performed by individually comparing the key point feature vectors of the images in *I* and the experiment dataset, and finding candidate matching features based on euclidean distance of their feature vectors. The percentage of the matched keypoints are calculated, i.e., the number of matching keypoints (Number of mKP) divided by the total number of keypoints (Total number of KP) for each image to condense the matching keypoints into a scalar quantity [16], [31]. Images are measured as similar if the percentage of similarity is above 80 percent. GCV is implemented by extracting the top five image labels and web entities. The relevance of the images is calculated from tags and entities with a similarity score above 0.5. We calculate the percentage of true positives to evaluate the accuracy and relevance of the service composition. GCV utilizes powerful pre-trained machine learning models through REST and RPC APIs. GCV API assigns labels to images and quickly classifies them into millions of predefined categories. Relevant images are selected based on the objects and spatiao-temporal metadata detected from similar images available over the Internet. Finally, we compare the composition performance of the approaches with our previous selection approach [5].

2. https://cloud.google.com/vision/, 2017 release

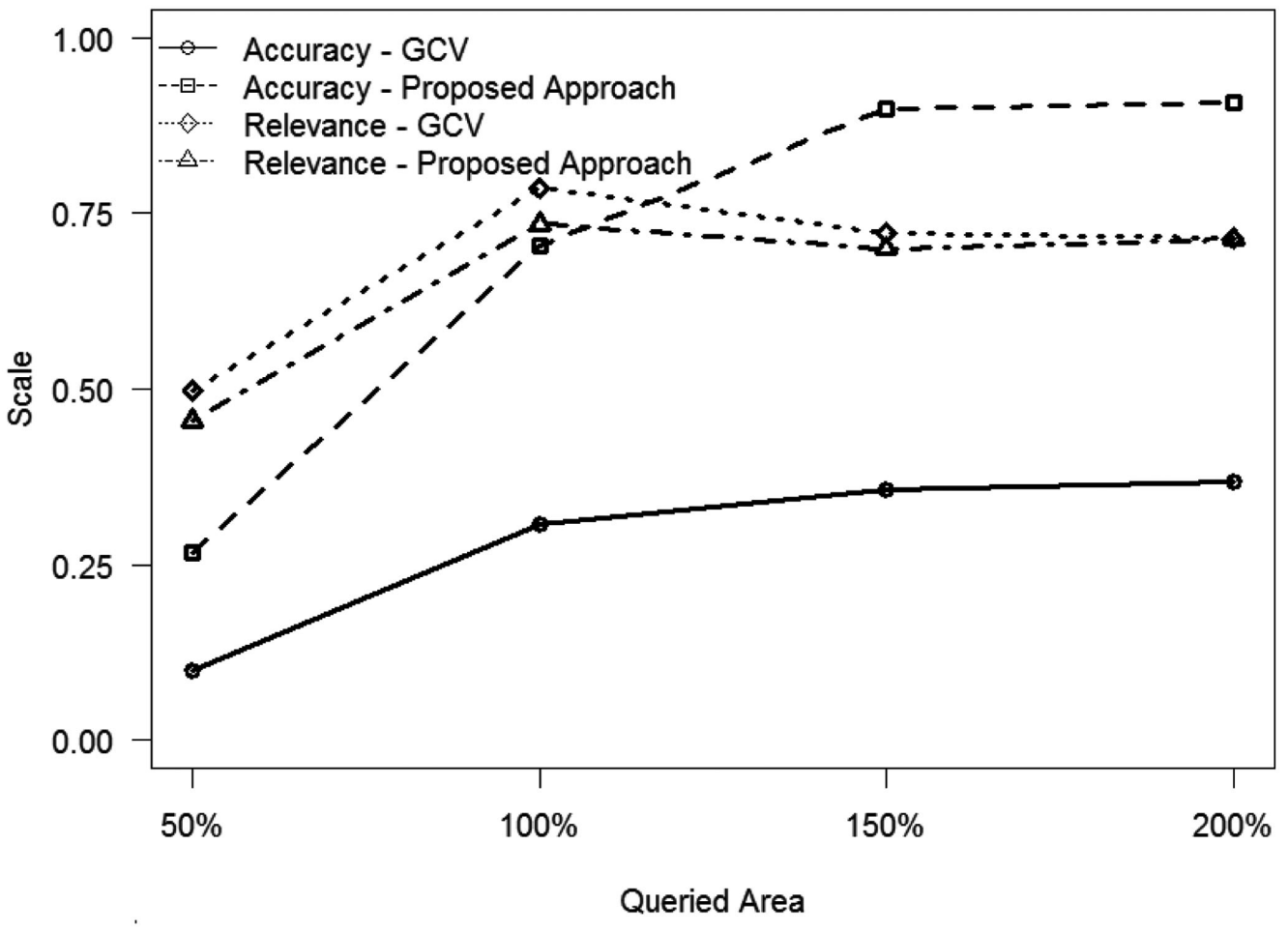

Fig. 10. Impact of the bounded region *BR* on accuracy and relevance.

The result depicts that the proposed approach shows significantly better performance than GCV, KAZE and SIFT. We observe that approximately 69 percent of services in the compositions are accurately categorised. The 31 percent error rate is reasonable due to the noises in the images. The proposed approach considers spatio-temporal parameters in addition to the service context. In contrast, GCV concentrates on the content of images and KAZE and SIFT focus on the visual features of the locations captured in the service. SIFT disregards the temporal and contextual bounds. Next, we assess the composition by analysing relevance (Table 9). The average relevance is 80 percent for the proposed approach and 79 percent for GCV. The performance of the selection approach is close to the proposed approach because both the approaches rely heavily on the context of the services and their relevance to the queries. The proposed approach shows the highest harmonic mean (Table 9). Therefore, it is concluded that the proposed approach outperforms the existing approaches on accuracy and relevance.

### 6.3.2 Scalability of Composite Services

The composition scalability indicates whether the proposed approach is feasible for larger compositions in terms of higher numbers of services available. As the number of services increases, the complexity of the proposed approach increases. Therefore, it is also important to evaluate the scalability in terms of the execution time of the composition with different sets of spatial parameters.

Two sets of experiments are conducted. First, we evaluate the performance of the proposed approach in terms of the composition *effectiveness* by expanding the query area. We evaluate the *relevance* and *accuracy* of the proposed approach and GCV on different query areas. SIFT and KAZE are not included in this experiment because they base on visual features of landmarks rather than spatial areas and locations of services. We assess how relevance and accuracy vary with 50, 100, 150 and 200 percent of the bounded region *BR* of a query. The number of services available also increases with the expansion of *BR*. The result indicates that the proposed approach produces higher accuracy in the composition. The result (Fig. 10) shows the impact of the *BR* on the accuracy and relevance of service composition. The result indicates that the proposed approach gives better results for smaller *BR* because the proposed approach focuses on event-based service selection using spatio-temporal parameters and service context. In contrast, GCV focuses on the content in the images rather than pinpointing the exact locations in the required time frame. The accuracy of the selected services in the composition is also higher for larger *BR* because the proposed approach is temporally and contextually more sensitive. Therefore, the proposed approach can ignore the noise services as the number of services increase, whereas GCV does not consider the increase in noise due to temporal bounds. Moreover, the accuracy of the selected services in the composition increases as the *BR* increases.

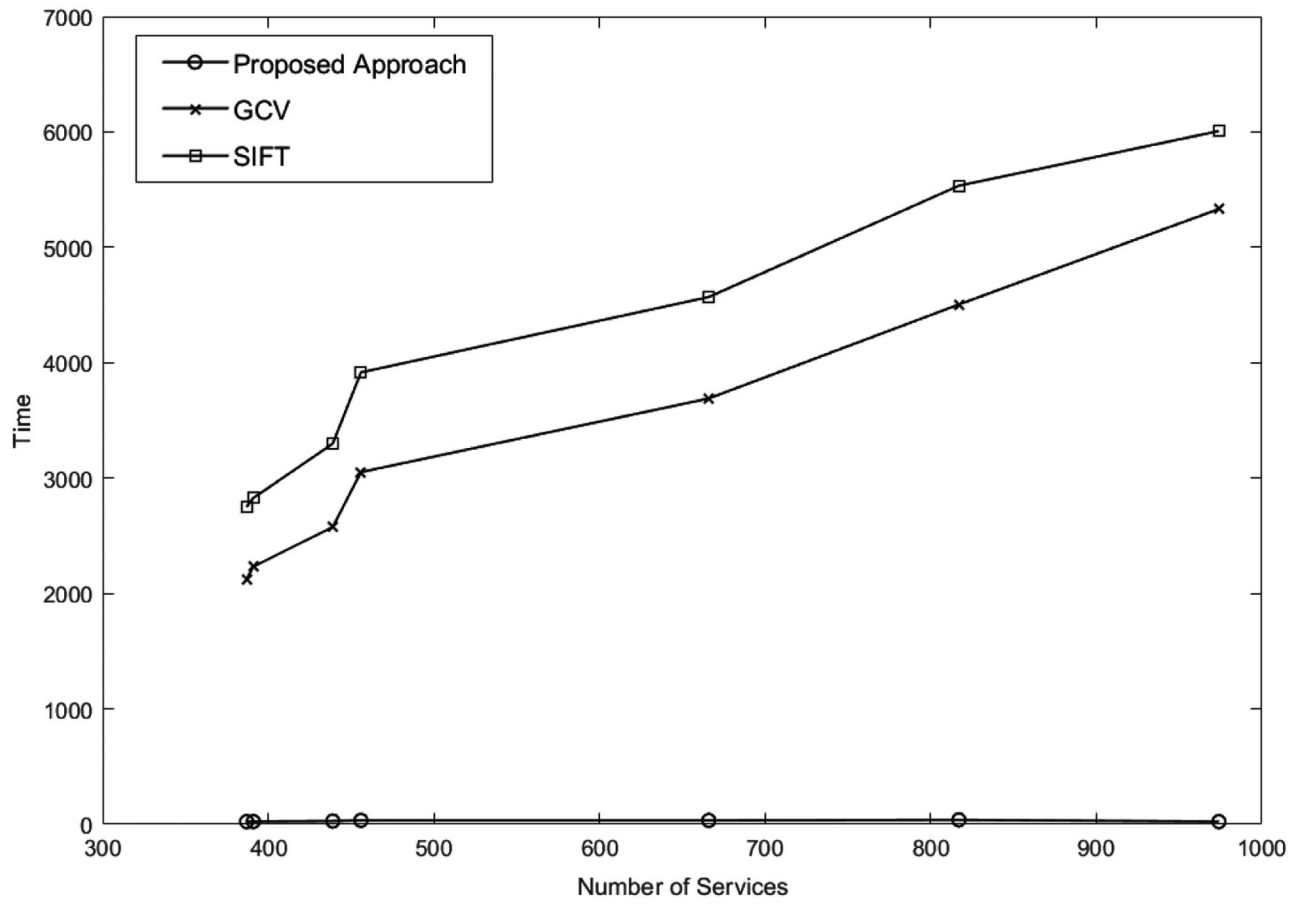

Fig. 11. Impact of the number of services on execution time.

After a specific variation, for example, 150 percent of the *BR* for the proposed approach, the accuracy rate becomes stable (Fig. 10). Whereas, GCV returns slightly higher relevance for the similar number of the selected relevant services (Fig. 10). As the area increases, the proposed approach retrieves almost the same number of relevant services of GCV. The result also concludes that loosening the spatial constraint results in similar results.

Next, we evaluate the scalability in terms of the impact of the number of services on the *execution time* of the composition. We investigate how the execution time varies with different numbers of services. This experiment depicts a comparison of execution time between three types of approaches: 1) SIFT, 2) GCV, and 3) the proposed approach. Given the set of 7 queries with different service densities in the query area, we calculate the average execution time. The number of services varies between 380 to 980, with different density within the query area. Fig. 11 shows the execution time of the three algorithms at different numbers of services.

The result shows that the proposed composition outperforms GCV and SIFT in terms of execution time (i.e., for around 500 services, the processing time is 30 ms versus 3,000 ms versus 4,000 ms). It is observed that the proposed approach takes a similar computation time, regardless of the number of services. This is because the non-relevant spatio-temporal services are filtered out in the initial steps, and a limited number of services require textual processing. The slight fluctuation shows the relative stability of the proposed approach.

TABLE 10
Human Perception Comparison

| Approach | SIFT | KAZE | GCV | Selection Approach | Proposed Approach |
|---|---|---|---|---|---|
| **Users' Selection** | 5% | 3% | 16% | 19% | 56% |

### 6.3.3 Human Perception Analysis of Composite Services

We conduct a real-world user survey to analyze users' perception of the performance of the proposed approach. We use MTurks[3] to generate an online survey. This survey is to assess which candidate approach provides the best visual summary for the seven event queries (Table 7). For all the queries we get the resultant visual summary from the five candidate approaches 1) SIFT, 2) KAZE, 3) GCV, 4) the selection approach, and 5) the proposed approach. The resultant output from these five approaches is evaluated by 1,100 MTurk users.

The result (Table 10) shows that 56 percent of users believe that the proposed solution provides the best visual summary of the image, i.e., the proposed approach can better capture and describe the queried events than the other approaches. The additional information regarding the location and timeline of the queried events provisioned in the product of the proposed approach is believed to be the main reason for users' high selection rate.

## 7 Conclusion

This paper proposes a SocSen service composition approach based on spatio-temporal and contextual proximity. The proposed approach leverages *crowdsourced* social media images to reconstruct tapestry scenes and provide related information for scene analysis. The major contribution is to deliver a unique SocSen service clustering and recommendation approach that considers 1) temporal-semantic relevance of the services to a queried scene, 2) spatio-temporal evolution of a scene, and 3) direction-oriented spatio-temporal composition to form a tapestry scene. We conducted experiments to evaluate the performance of the proposed approach for accurate and effective composition.

In the future, we plan to focus on the specific spatial aspects (i.e., angle and direction) of SocSen service composition and conducting experiments on social media video clips. In addition, we plan to improve the smoothness of tapestry scenes by metricizing the overlaps and coverage of the SocSen services in a composition.

## Acknowledgments

This work was supported by NPRP9-224-1-049 Grant from the Qatar National Research Fund (a member of Qatar Foundation) and DP160100149 and LE180100158 Grants from Australian Research Council. The statements made herein are solely the responsibility of the authors.

3. https://www.mturk.com/

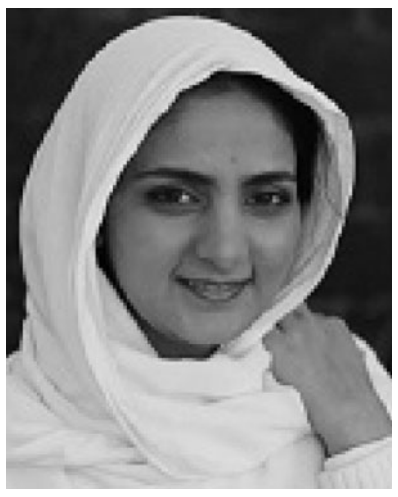

**Tooba Aamir** received the BS degree in computer engineering from COMSATS University, Islamabad, Pakistan, in 2010, and the MS degree in software systems engineering from the University of Melbourne, Melbourne, Australia, in 2012. She is currently working toward the PhD degree in the School of Science, RMIT University, Melbourne, Australia. Her research interests include social-sensor cloud service modelling, selection and composition for scene reconstruction, and analysis.

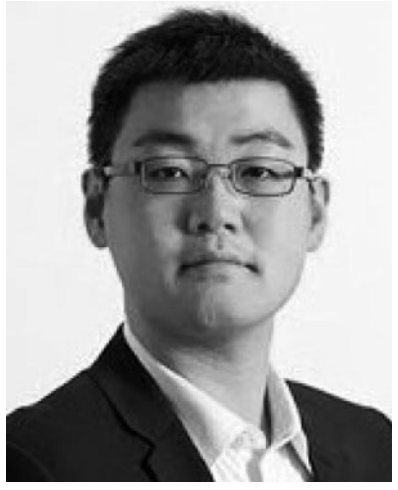

**Hai Dong** (Member, IEEE) received the PhD degree from the Curtin University of Technology, Perth, Australia. He is currently a lecturer with the School of Science, RMIT University, Melbourne, Australia. He was previously a vice-chancellor's research fellow with RMIT University. He has published a monograph and more than 80 research publications in international journals and conferences, such as the *Communications of the ACM*, *IEEE Transactions on Services Computing*, *IEEE Transactions on Industrial Informatics*, *IEEE Transactions on Industrial Electronics*, *Journal of Computer and System Sciences*, *World Wide Web Journal*, ICSOC, ICWS, etc. He received the Best Paper Award in ICSOC 2016. His primary research interests include: services computing, distributed systems, cyber security, artificial intelligence, and data mining. He is a member of the ACM.

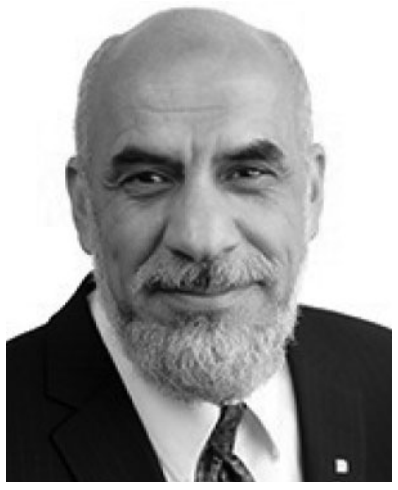

**Athman Bouguettaya** (Fellow, IEEE) received the PhD degree in computer science from the University of Colorado at Boulder, Boulder, Colorado, in 1992. He is currently a professor and head of the School of Computer Science, University of Sydney, Australia. He was previously science leader in service computing with CSIRO ICT Centre, Canberra, Australia. Before that, he was a tenured faculty member and program director with the Computer Science Department, Virginia Tech. He is or has been on the editorial boards of several leading journals including, the *IEEE Transactions on Services Computing*, *IEEE Transactions on Knowledge and Data Engineering*, *ACM Transactions on Internet Technology*, *ACM Computing Surveys*, and *VLDB Journal*. He has published more than 250 books, book chapters, and articles in journals and conferences in the area of databases and service computing (e.g., the *IEEE Transactions on Knowledge and Data Engineering*, *ACM Transactions on the Web*, *World Wide Web Journal*, *VLDB Journal*, SIGMOD, ICDE, VLDB, and EDBT). He was the recipient of several federally competitive Grants in Australia (e.g., ARC) and the US (e.g., NSF, NIH). He is a distinguished scientist of the ACM.